\documentclass{SCGE}

\begin{document}

%%%%%%�°�ʽҪ��������
\begin{picture}(0,0){\rm
\put(0,-20){\makebox[160truemm][l]{\bf {\sanhao\raisebox{2pt}{.}}
Invited Review  {\sanhao\raisebox{1.5pt}{.}}}}}
\put(0,-34){\jiuwuhao {\textcolor[rgb]{0.5,0.5,0.5}{\sf %Special Topic: Fluid Mechanics
}}}%%(11��ע�ͣ���\textcolor[rgb]{x,x,x}�е�����xԽ��Խ��)
\end{picture}

\def\bm{\boldsymbol}

\def\dl{\displaystyle}
\def\tlj{\end{document}}
\def\d{{\rm d}}
\def\e{{\rm e}}
\def\i{{\rm i}}

% The author doesn't need fill in it.
\Year{2015} %
\Month{January} %
\Vol{58} %  ����
\No{1} %  �ں�
\BeginPage{1} % ��ҳ��
\AuthorMark{{\rm Low B. C.}}  %(11��ע�ͣ�ҳü�ϵ�����)
\AuthorMarkCite{{\rm Low B C.}~} %(11��ע�ͣ�citation�е�����)
\DOI{10.1007/s11433-014-5626-7} % The author doesn't need fill in it.
\ArtNo{015201}

% \title[short text for running head]{full title}{comments for title}
\title[Field topologies in ideal and near-ideal magnetohydrodynamics and vortex dynamics]
{Field topologies in ideal and near-ideal magnetohydrodynamics \\ and vortex dynamics$^\dag$}%���� {�к�����ʱ����Ҫ�����⸴������������棬ʹ����������ʾ���塣û�к�����ʱ������ſ���ɾ��}

\author[*]{LOW B. C.}{}
%\footnote{*Corresponding author (email: dlzhangyu@yahoo.com.cn)}%�ֶ�E-mail��ַ

\address[]{High Altitude Observatory, National Center for Atmospheric Research,
Boulder, CO 80307, USA}

\maketitle \vspace{-3.5mm}{\footnotesize\begin{center} Received October 21, 2014; accepted October 30, 2014; published online December 1, 2014%�ո�����
\end{center}}\vspace*{-5mm}

%     Abstract is required.
\begin{center}
\rule{16.5cm}{0.4pt}
\parbox{16.5cm}
{\begin{abstract} Magnetic field topology frozen in ideal magnetohydrodynamics (MHD) and its breakage in near-ideal
MHD are reviewed in two parts, clarifying and expanding basic concepts.  The first part gives a physically
complete description of the frozen field topology derived from magnetic flux conservation as the
fundamental property, treating four conceptually related topics: Eulerian and Lagrangian descriptions
of three dimensional (3D) MHD, Chandrasekhar-Kendall and Euler-potential field representations,
magnetic helicity, and inviscid vortex dynamics as a fluid system in physical contrast to ideal MHD.
A corollary of these developments clarifies the challenge of achieving a high degree of the frozen-in
condition in numerical MHD. The second part treats field-topology breakage centered around the Parker
Magnetostatic Theorem on a general incompatibility of a continuous magnetic field with the dual demand
of force-free equilibrium and an arbitrarily prescribed, 3D field topology.  Preserving field topology
as a global constraint readily results in formation of tangential magnetic discontinuities, or,
equivalently, electric current-sheets of zero thickness.  A similar incompatibility is present
in the steady force-thermal balance of a heated radiating fluid subject to an anisotropic
thermal flux conducted strictly along its frozen-in magnetic field in the low-$\beta$ limit.
In a weakly resistive fluid the thinning of current sheets by these general incompatibilities
inevitably results in sheet dissipation, resistive heating and topological changes in the
field notwithstanding the small resistivity.  Strong Faraday induction drives but
also macroscopically limits this mode of energy dissipation, trapping or storing free energy in self-organized ideal-MHD structures.  This
property of MHD turbulence captured by the Taylor hypothesis is reviewed in relation to the Sun's corona, calling for a basic quantitative description of the breakdown of flux conservation in the low-resistivity limit.  A cylindrical initial-boundary value problem provides specificity in the general MHD ideas presented.
\end{abstract}}
\end{center}\vspace*{-0.6cm}

\begin{center}
\parbox{16.5cm}
{\bf\jiuhao magnetic topology, magnetic reconnection, current sheets, magnetic helicity,
thermal conduction, solar corona, magnetohydrodynamics, interstellar clouds}%�ؼ���
\end{center}

\begin{center}
{}%������
\CITA    %%(11��ע�ͣ�Citation�����Զ�����)
%\Cit{~~~???, et al. ???. Sci China-Phys Mech Astron, 2014, 57: 1--6, doi:}%%(11��ע�ͣ�Citation�������ֶ���д)
\end{center}

\textwidth=178truemm \textheight=236truemm%%%%%%�°�ʽҪ����

%%%%%%%%%%%%%%%%%%%%%%%%%%%%%%%%%%%%%%%%%%%%%%%%%%%%%%%%%%%%
\wuhao\vspace*{1.5mm}

\begin{multicols}{2}

%%%%%%%%%%%%%%%%%%%%%%%%%%%%%%%%%%%%%%%%%%%%%%%%%%%%%%%%%%%%
%% Text of article.
%%%%%%%%%%%%%%%%%%%%%%%%%%%%%%%%%%%%%%%%%%%%%%%%%%%%%%%%%%%%
%    Section headings
\renewcommand{\baselinestretch}{1.08} \baselineskip 12.2pt\parindent=10.8pt

\renewcommand{\thefootnote}

\section{Introduction}

Magnetohydrodynamics (MHD) describes plasmas as an electrical fluid conductor.  The ideal fluid conductor is singular as the limiting case of highly conducting fluids.  Recent investigations of the topological properties of magnetic fields

\noindent\rule{2.5cm}{0.4pt}\\[0.1mm]{\qihao *Corresponding author (email:
low@ucar.edu)\vspace{-1mm}\\
\dag Recommended by CHEN PengFei (Associate Editor)}%�ֶ�E-mail��ַ

\noindent  in Newtonian MHD have clarified and expanded our understanding of this singular limit.  We review these developments based on the dissipative MHD equations for a single fluid:
\begin{align*}
 \rho \left[\frac{\partial {\bm v}}{\partial t} + ({\bm v} \cdot \nabla) {\bm v} \right] =\;&\frac{1}{4 \uppi}(\nabla \times {\bm B}) \times {\bm B} - \nabla p \\
\;&+ \nu_1 \nabla^2 {\bm v}  + \nu_2 \nabla (\nabla \cdot {\bm v}), \tag 1\vspace*{-3mm}
\end{align*}
$$\frac{\partial {\bm B}}{\partial t} = \nabla \times( {\bm v} \times {\bm B}) + \eta \nabla^2 {\bm B},\eqno (2)$$
$$\frac{\partial \rho}{\partial t} + \nabla \cdot (\rho {\bm v})=0,\eqno (3)
$$
describing a magnetic field ${\bm B}$ in a fluid of density $\rho$ and pressure $p$ moving with velocity ${\bm v}$.
Of the great complexity of particle-particle and particle-field processes in
a plasma [1--5], we retain only viscosity and Ohmic resistivity described in their simplest forms by the constant coefficients $(\nu_1, \nu_2, \eta)$.  In the CGS units used in this paper, $\eta=\frac{c^2}{4 \uppi \sigma}$, where $c$ is the speed of light and $\sigma$ is a constant Ohmic conductivity, and properties will be discussed equivalently in terms of the conductivity or resistivity.  For brevity, we refer to ${\bm B}$ simply as the field.  The solenoidal condition $\nabla \cdot {\bm B}=0$ is implied by the resistive induction equation (2).

The electric field ${\bm E}$ in the laboratory frame is a derived quantity in MHD, given by Ohm's law\vspace{-2mm}
$$
{\bm E} + \frac{1}{c} {\bm v} \times {\bm B} = \frac{\eta}{c} \nabla \times {\bm B},\eqno (4)\vspace{-2mm}
$$
where on the left side is the electric field in the rest frame of an infinitesimal parcel of fluid.
 Charge separation and the electric force are negligible in the sub-relativistic momentum equation
 (1) whereas the field exerts a Lorentz force on the fluid.  The inductive
effect of the flowing conducting fluid together with Ohmic dissipation influence the field under Faraday's law of induction equation (2).  Introducing the temperature $T$ of the fluid by an equation of state, the system of equations can then be closed by mass conservation equation (3) plus an energy-transport equation, yet to be specified, to determine the variables $(p, \rho, T, {\bm v}, {\bm B})$.

Our interest is focused on the fluid and field behaviors in the regime of high conductivity, i.e., $\sigma \rightarrow \infty$, or, equivalently, the regime of low resistivity $\eta \rightarrow 0$.  To that end we carry out two tasks in our review, to construct a physically complete description of the field in an $\eta=0$ ideal fluid conductor and, based on the description as a reference, to understand the nonlinear couplings in the regime $\eta \rightarrow 0$ among Faraday induction, the dynamical forces and energy transport.

The ideal conductor is described by\vspace{-2mm}
$$\frac{\partial {\bm B}}{\partial t} = \nabla \times ({\bm v} \times {\bm B}),\eqno (5)\vspace{-2mm}
$$
setting $\eta=0$ in eq. (2), with the electric field given by\vspace{-2mm}
$$
{\bm E} =- \frac{1}{c} {\bm v} \times {\bm B},\eqno (6)\vspace{-2mm}
$$
due entirely to Faraday induction.  The conservation of the magnetic flux across {\it every} fluid surface in an MHD evolution is the defining property of the ideal induction equation.  Derived from it is the well known property that the evolving field preserves its topology.  The field-fluid interaction at each point in space subject to this global topological constraint can readily produce tangential discontinuities (TDs) in the field [6--24]. TDs contain unbounded but integrable electrical current densities, i.e., current sheets (CSs) of zero thickness, and we shall discuss physical issues in terms of TDs or CSs interchangeably. Such current singularities are physically admissible in the complete absence of resistivity.

A highly conducting fluid with a weak $\eta \ne 0$ naturally also conserves magnetic flux with the same
tendency to form TDs but only on length scales much larger than a lower bound fixed by $\eta$.
A steepening field gradient characterized by some monotonically-decreasing small scale $l$ eventually
and inevitably undergoes resistive dissipation at a sufficiently short diffusion time-scale $t_{\rm d}=\frac{l^2}{\eta}$
relevant to dynamics.  Field topology ceases to be preserved on the large scale in consequence of resistive magnetic reconnection on the small scales [5,8,25--28].  Thus field steepening by high conductivity leads to resistive change in field topology and the highly conducting fluid is distinct from the $\eta=0$ ideal fluid.  If a gas is hot enough to be fully ionized, such a highly conducting fluid may sustain its high temperature via the spontaneous formation and resistive dissipation of CSs, an attractive explanation for the million-degree hot coronae of the Sun and billions of solar-like stars in our Galaxy [7,8,29].

The $\eta=0$ ideal fluid being a singular limit corresponds to the removal from Faraday's law of induction equation (2) of its highest-order spatial differential-operator that describes the resistive diffusion of the field.  In classical hydrodynamics, the regime of turbulent phenomena corresponds to the coefficients of viscosity $(\nu_1, \nu_2)$ being sufficiently small by some measure [30].  The complete removal of the diffusion operator on ${\bm v}$ from momentum equation (1) with $(\nu_1, \nu_2) \equiv 0$ is similarly a singular limit. Thus we also expect turbulent field behaviors when $\eta$ is sufficiently small.  Recent theoretical studies [16,31] have shown that the degree of complexity in the time-dependent MHD solutions may increase without bound, by some suitable measure, as $\eta \rightarrow 0$.  Our review avoids such formidable fundamental problems but will in its course encounter this turbulent nature of high-conductivity.

In sect. 2 a physically complete description of the field frozen in an ideal fluid is constructed,
treating in a logical sequence magnetic flux conservation as a fundamental property,
Lagrangian and Eulerian descriptions [32,33], Chandrasekhar-Kendall [33--35] and Euler-potential [32,33,36]
field representations, magnetic helicity [33,37--41] and, inviscid vortex dynamics in an incompressible
neutral fluid in physical contrast to ideal MHD. In sect. 3 an overview is given on the breaking of
field topologies in a weakly resistive fluid via spontaneous CSs, treating the Parker Magnetostatic
Theorem [8,10,17,18], the Taylor hypothesis [35,42--46] and and its generalization, and a CS-producing coupling between MHD and anisotropic thermal conduction [19].  The solar corona [13,14,47--50] is briefly described to provide physical context and motivation for the developments reviewed. The review takes the approach of first clarifying a physical idea and then seeking mathematical rigor whenever possible in its description. Sect. 4 gives a summary and discussion.\vspace{-1mm}

\section{Frozen field topologies in ideal fluids}\vspace{-1mm}

Rewrite ideal induction equation (5) for a perfect fluid in the

\noindent form
$$
\frac{\partial {\bm B}}{\partial t} + ({\bm v} \cdot \nabla) {\bm B}= ({\bm B} \cdot \nabla) {\bm v}-(\nabla \cdot {\bm v}) {\bm B}.\eqno (7)
$$
The left side is the rate of change of ${\bm B}$ observed by a fluid particle moving with velocity ${\bm v}$.  The change is due to a stretching and compression of the field described by the two terms on the right side.  The induction equation describes the conservation of magnetic flux:
$$
\frac{{\rm d} F}{{\rm d}t} \equiv \frac{{\rm d}}{{\rm d}t} \int_{\Sigma(t)} {\bm B} \cdot {\rm d}{\bm S} = \frac{{\rm d}}{{\rm d}t} \int_{\gamma(t)} {\bm A} \cdot {\rm d}{\bm s}=0,\eqno (8)
$$
where $F$ is the total magnetic flux across any {\it fluid surface} $\Sigma$, with oriented area element ${\rm d}{\bm S}$, evolving in the velocity field ${\bm v}$.  The above alternative statement of flux conservation expresses $F$ as a line integral of the vector potential ${\bm A}$ along the closed boundary $\gamma$ of $\Sigma$, with directed path element ${\rm d}{\bm s}$, where
$$
{\bm B}  = \nabla \times {\bm A}.\eqno (9)
$$
Unless otherwise stated we treat simply-connected closed curves, surfaces and volumes.  For a given ${\bm B}$, its vector-potential ${\bm A}$ is not unique up to a free gauge $G$, so every magnetic property expressed in terms of ${\bm A}$ must be rendered gauge-independent.  That is, the property must be invariant under the transformation ${\bm A} \rightarrow {\bm A}+\nabla G$ for any $G$ which is the case for $F$.

Magnetic flux $F$ is not a differential property at a point in space but an integral over a 2D surface.  A field line of a given field ${\bm B}$ described by the ordinary differential equations (ODEs):
$$
\frac{{\rm d}x}{B_x}=\frac{{\rm d}y}{B_y}=\frac{{\rm d}z}{B_z},\eqno (10)
$$
in Cartesian coordinates $(x, y, z)$ carries no flux.  It is a common notion that a single field line represents a thin tube of flux with the field line passing through the small cross-section of the tube.  Let us put aside this notion until we return to it in sect. 2.3.  The conservation law on $F$ over any fluid surface makes a basic point concerning whether to treat a fluid in terms of its local properties at each point in space or its globally defined properties.  Among the latter properties are those of a topological nature.

A geometric object is defined by the metric of distance in 3D Euclidean space.  Such an object has topological properties independent of the metric that remain meaningful when the object is subject to {\it all continuous} deformations.  For example, consider two closed tubes of fluid identified at a given time to be linked an integer $\pm N$ times about each other, signed according to whether the link is right or left handed, respectively.  This linkage is a global property invariant as the two fluid tubes evolve in a continuous flow.  Global properties are expressed by integral equations whereas the PDEs describe local conditions.

MHD eqs. (1)--(3) are an Eulerian description in terms of functions of space-time with no interest in knowing where each fluid particle is located at a given time.  This description is traditionally the preferred one because of its intrinsic simplicity, especially in numerical computations.   The Lagrangian description identifies specific fluid parcels to be followed in the course of their motions and mutual interactions.  The fluid parcels do not have to be point like.  The nature of an investigation may conceivably motivate adopting a Lagrangian description that partitions a fluid into components to be followed during an evolution.  Such a description is put to good use in this section, recognizing the following essential point of the review.   Flux conservation is the fundamental physical property from which all $\eta=0$ field properties are derived and this conservation law is a Lagrangian statement.

\subsection{A cylindrical initial-boundary value problem}

The following initial-boundary value problem provides physical specificity in the general ideas to be discussed.  Consider a field ${\bm B}$ in a perfect fluid filling the upright cylindrical domain $V\!: R<R_0, |z|<z_0$, of length $2 z_0$ and radius $R_0$, using cylindrical coordinates $(R, \varphi, z)$.  The field satisfies the boundary conditions:
\begin{align*}
\;& R=R_0, \quad B_R=0,\tag {11}\\
\;& z=\pm z_0, \quad B_z=b_{\pm}(R, \varphi),\tag {12}
\end{align*}
where $b_{\pm}(R, \varphi)$ are prescribed.  Since the field is tangential at the cylindrical side of $V$, the boundary-flux distributions at $z=\pm z_0$ are subject to the solenoidal condition:
\begin{align*}
\int_{0}^{R_0}~\int_0^{2\uppi}~b_{+}(R, \varphi) ~ {\rm d} \varphi R {\rm d}R =\;& \int_{0}^{R_0}~\int_0^{2\uppi}~b_{-}(R, \varphi) ~ {\rm d} \varphi R {\rm d}R\\
 =\;& F_0,\tag {13}
\end{align*}
allowing for a constant net axial flux $F_0$ along $V$. There are two classes of fields in $V$, the wholly contained fields with $b_{\pm}(R, \varphi) \equiv 0$, for which $F_0=0$, and the anchored fields with $b_{\pm}(R, \varphi) \ne 0$, for which both cases of $F_0=0$ and $F_0 \ne 0$ are admissible.

For simplicity suppose the wall at boundary $\partial V$ is a rigid perfect conductor.   Unless stated otherwise let the fluid be inviscid, i.e., $\nu_1=\nu_2=0$ in momentum equation (1).  The momentum equation then imposes no condition on the tangential boundary velocity at $\partial V$.  For the contained field, with  $b_{\pm}(R, \varphi) \equiv 0$, induction equation (5) also imposes no condition on the tangential velocity at $\partial V$.  For the anchored field, $b_{\pm}(R, \varphi) \ne 0$,  the induction equation imposes no condition on the tangential velocity at $R=R_0$ but demands that the velocity vanishes at $z=\pm z_0$.  These boundary conditions ensure the electromagnetic condition that electric field ${\bm E}$ given by eq. (6) is tangentially continuous across $\partial V$ into ${\bm E}=0$ in the boundary wall.  We summarize these boundary conditions:
$$\begin{array}{ll}
b_{\pm}(R, \varphi) \equiv 0 \Rightarrow R=R_0, ~ v_R=0;~ z=\pm z_0, ~ v_z=0, \\
b_{\pm}(R, \varphi) \ne 0 \Rightarrow R=R_0, ~ v_R=0;~ z=\pm z_0, ~ {\bm v}=0,
\end{array}\eqno (14)$$
to be referred to as the electromagnetic conditions on the velocity.  If friction is present with $(\nu_1,\nu_2) \ne 0$, then the momentum equation independently imposes the boundary conditions:
$$
{\rm at}~R=R_0~{\rm and}~z=\pm z_0, ~~ {\bm v}=0,\eqno (15)
$$
which take precedence over the electromagnetic conditions.

\subsection{A toroidal Lagrangian description of MHD}

Consider the partition [32,33] of a given fluid in $V$ into an exhaustive set of disjoint contiguous toroids denoted by $\tau$.  This construction is quite independent of whether the fluid has a magnetic field.  To keep the partition topologically elementary the $\tau$ toroids are constructed unlinked in the sense that any number of them can be continuously deformed to separate one from another without entanglement.

Consider a continuous field ${\bm B}_{\tau}$ in $V$ with field lines that are unlinked closed curves in $V$.  The field thus comprises unlinked closed tubes of untwisted flux that serve as a particular realization of the $\tau$ toroid partition of the fluid.  We shall refer to the fluid and field in terms of their $\tau$ toroids (of fluid) and $\tau$ flux-tubes, respectively, keeping the two partitions conceptually separate for a reason that will become clear.

Once the common partition of field and fluid defined by ${\bm B}_{\tau}$ is given at any initial time, it is permanently
identified for all time  under flux conservation.  Initially the boundary of each $\tau$ toroid is a flux surface, across each sub-area of which there is zero flux.  Along the interior of each $\tau$ toroid is a constant net axial flux $f(\tau)$.  In subsequent evolution flux conservation ensures that the
fluid boundary of a $\tau$-toroid of fluid remains a flux surface bounding a flux tube with the unchanging axial flux $f(\tau)$.   The care taken in this description recognizes that only the fluid has identity whereas a flux tube under flux conservation acquires its identity by the fluid in which it is permanently embedded.

Let each $\tau$ toroid have an infinitesimally narrow cross-section $d\sigma(\tau){\hat l}$ varying along the toroid; there is an enormously large total number of $\tau$ toroids.  Then at any time by locating each of these $\tau$ toroids we can construct the evolved field ${\bm B}_{\tau}$ as a function of space from the conserved flux:
$$
f(\tau)={\bm B}_{\tau} \cdot d\sigma(\tau){\hat l},\eqno (16)
$$
to any desired precision, ${\hat l}$ being the unit vector along a narrow $\tau$ toroid.

Now consider a more complicated field obtained by the linear superposition ${\bm B}={\bm B}_{\tau}+{\bm B}_{\upsilon}$ where ${\bm B}_{\upsilon}$ is an independently prescribed field satisfying boundary conditions (11) and (12).   Suppose $({\bm B}_{\tau}, {\bm B}_{\upsilon})$ each satisfies induction equation (5) for the same fluid velocity ${\bm v}$.  In addition to the constants of motion $f(\tau)$, we also have the constants of motion
$$
F(\tau, \upsilon)=\int_{\Sigma(\tau)}~ {\bm B} \cdot {\rm d}{\bm S}= \int_{\Sigma(\tau)}~ {\bm B}_{\upsilon} \cdot {\rm d}{\bm S}\eqno (17)$$
where $\Sigma(\tau)$ is any fluid surface subtended by the closed curve represented by a $\tau$ toroid.  That is, $\Sigma(\tau)$ is just a geometric surface closing the hole of the toroid.  The total flux $F(\tau, \upsilon)$ passing through that hole is contributed entirely by ${\bm B}_{\upsilon}$ because ${\bm B}_{\tau}$ makes no contribution.  For all fields of the form ${\bm B}={\bm B}_{\tau}+{\bm B}_{\upsilon}$, the toroidal partition of the fluid based on ${\bm B}_{\tau}$ yields two sets of Lagrangian constants of motion, namely, $[ f(\tau), F(\tau, \upsilon)]$.  It follows that the summation
$$
{\mathcal H}=\sum_{\tau} {\mathcal F}_{\tau} \left[ f(\tau), F(\tau, \upsilon) \right],\eqno (18)
$$
to be referred as the general Lagrangian helicity, is also a constant of motion, where ${\mathcal F}_{\tau}$ is any prescribed function of two variables for each $\tau$.  The arbitrary function ${\mathcal F}_{\tau}$ generates a two-dimensional continuum of conserved values of ${\mathcal H}$ as an expression of the fact that magnetic flux is conserved on all fluid surfaces in a flow.

Consider the special case of ${\mathcal H}$ as a sum of simple products:
$$H_{\rm L}= \sum_{\tau}  f(\tau) F(\tau)= \sum_{\tau}  {\bm B}_{\tau} \cdot d\sigma(\tau){\hat l} \int_{\Sigma(\tau)}~ {\bm B}_{\upsilon} \cdot {\rm d}{\bm S}.\eqno (19)
$$
Denote by $\gamma(\tau)$ the closed path defined by the toroid $\tau$ of infinitesimal cross-section $d\sigma(\tau)$ and express the solenoidal ${\bm B}_{\upsilon}=\nabla \times {\bm A}_{\upsilon}$ in terms of its vector potential.  One application of Stokes law gives
\begin{align*}
H_{\rm L}=\;& \sum_{\tau}  {\bm B}_{\tau} \cdot d\sigma(\tau){\hat l} \oint_{\gamma(\tau)}~ {\bm A}_{\upsilon} \cdot {\rm d}{\bm l} \\
=\;& \sum_{\tau}  {\bm A}_{\upsilon} \cdot {\bm B}_{\tau} d\sigma(\tau){\rm d}l\\
\label{H_L}
=\;& \int_V~ {\bm A}_{\upsilon} \cdot {\bm B}_{\tau} ~{\rm d}V,\tag {20}
\end{align*}
writing ${\rm d}V=d\sigma(\tau){\rm d}l$ in the limit of an infinitesimal $d\sigma(\tau)$.  In that limit, ${\bm B}_{\tau}$ is directed in the direction ${\hat l}$ along each toroid and $d\sigma(\tau){\rm d}l$ defines a differential volume so that summing over $\tau$ gives the volume integral obtained.  The helicity $H_{\rm L}$ is a constant of motion, an integral over the fluid volume $V$ involving $[{\bm A}_{\upsilon}, {\bm B}_{\tau}]$ as Eulerian variables at any given time $t$.  The Lagrangian nature of $H_{\rm L}$ is inseparable, its calculation requiring knowledge of all the $\tau$ toroids at time $t$, where each is located and how each has been deformed from its shape at initial time $t_0$.

The central point of the construction is that ${\bm B}$ in its topological complexity has been expressed in terms of two component fields $\left({\bm B}_{\tau}, {\bm B}_{\upsilon}\right)$ that are topologically simpler.  Here we have shown that the Lagrangian statement of flux conservation has led naturally to the concept of a helicity $H_{\rm L}$ that is only a number in a continuum of constants of motion. We next show that such a linear decomposition can, in fact, be carried out for any prescribed ${\bm B}$ in $V$.

\subsection{Field representations by linear decomposition}

The simple prescription of ${\bm B}$ as a function of space hides a general topological complexity of its field lines and flux surfaces defined by the ODEs (10).  In the neighborhood of a chosen point in space, this pair of first order ODEs has a general solution
$$
[x, y]=[x(z, \xi_0, \zeta_0), y(z, \xi_0, \zeta_0)],\eqno (21)
$$
treating $z$ as the independent variable and introducing a pair of integration constants $(\xi_0, \zeta_0)$.  This solution describes the field lines in the neighborhood, identifying each field line by a pair of values of $(\xi_0, \zeta_0)$ determined from the coordinates of any particular point $(x_0, y_0, z_0)$ on the field line.  Visualize the two-parameter continuum of field lines so constructed in terms of two independent families of flux surfaces on which the field lines lie.  To describe these flux surfaces, treat eq. (21) as a pair that determine integration constants $(\xi_0, \zeta_0)$ as unknowns.  Denote the solution formally as:
$$
[\xi,  \zeta]=[\xi (x, y, z), \zeta(x, y, z)].\eqno (22)
$$
Each field line is then given by $(\xi, \zeta)=(\xi_0, \zeta_0)$ as the intersection between two level surfaces of constant $(\xi, \zeta)$, defining two independent families of flux surfaces.

\subsubsection{Euler potentials}

Flux surfaces can be used to express a solenoidal field in the form:
$$
{\bm B}=b(\xi, \zeta) \nabla \xi \times \nabla \zeta,\eqno (23)
$$
geometrically describing the field to be along field lines $(\xi, \zeta)=(\xi_0, \zeta_0)$ with a constant amplitude $b(\xi, \zeta)$ on each field line. The pair $(\xi, \zeta)$ called the Euler potentials [32,33,36] is not unique, for it depends on how the two constants of integration $(\xi_0, \zeta_0)$ reside in the actual calculated expression of the solution (21).  This non-uniqueness is not physically significant because it merely indicates that any given set of field lines may be ordered in an infinite number of ways into two independent families of flux surfaces.

The Euler-potential field representation has a fundamental difficulty.  Unlike fields with an ignorable coordinate, solution (21) for a 3D field generally exists only as a local solution around the chosen point without an assurance that the Euler potentials can be defined globally.

Fully 3D fields not anchored to the domain boundary generally may be ergodic [51,52], with field lines each of infinite length and filling up a finite-sized volume.  By integrating ODEs (10) far enough along such a field line will bring the line to as close to any point in the sub-volume as desired.  In the language of chaos dynamics [53], ODEs (10) are not integrable in the sense of a general absence of globally defined integrals.  If a field line is volume filling, so are any two surfaces intersecting along that line.  Then we have the geometric absurdity that every point in the 3D volume lies on these surfaces.  Yet solution (21) exists in any local neighborhood.  This mathematical trouble manifests itself via the Euler potentials $(\xi, \zeta)$ being necessarily multi-valued on global scales.

Anchored fields in the cylindrical domain $V$ may embed sub-systems of ergodic flux.  Anchored fields may also have field lines that diverge exponentially in a finite subdomain [8,11,15,23,53].  To be sure, non-ergodic fields constitute an infinite set, but this set by their nature is a subset of measure zero of the set of all field topologies admissible in $V$.

There are two resolutions to the difficulty of describing a field in terms of a pair of Euler potentials.   We may so represent a field in as many localities in space as needed but suitably connecting the distinct Euler-potential representations across the boundaries separating the localities, an unattractive formidable undertaking.  The other resolution [32] is attractive, which is to decompose the field into a linear sum
$${\bm B}=\sum_{i=1}^N ~{\bm B}_i=\sum_{i=1}^N ~b_i(\xi_i, \zeta_i) \nabla \xi_i \times \nabla \zeta_i,\eqno (24)
$$
of $N$ fields of simpler topologies, simplicity to be defined shortly, each field evolving according to induction equation (5) with the common fluid velocity governed by the momentum equation.  The Lorentz force is defined in terms of ${\bm B}$ as the linear sum of these fields.  The induction equation can then be integrated once with respect to space to give
$$\begin{array}{ll}
\dl\frac{\partial \xi_i}{\partial t}+ {\bm v} \cdot  \nabla \xi_i = 0,\vspace{2mm}\\
\dl\frac{\partial \zeta_i}{\partial t}+ {\bm v} \cdot  \nabla \zeta_i=0.
\end{array}\eqno (25)$$
Simpler topology is meant that each ${\bm B}_i$ has a global pair of single-valued Euler potentials.  Therefore, representation (24) is global.  Now we turn to an explicit decomposition of this kind for the cylindrical field.

\subsubsection{Chandrasekhar-Kendall representation}

Any solenoidal field ${\bm B}$ in $V$ can be expressed as the linear superposition of the two solenoidal fields:
\begin{align*}
{\bm B} =\;& {\bm B}_{\Phi} + {\bm B}_{\Psi}, \tag {26}\\
{\bm B}_{\Phi}=\;& \nabla \times \left( \Phi {\hat z} \right)=\frac{1}{R} \frac{\partial \Phi}{\partial \varphi} {\hat R} - \frac{\partial \Phi}{\partial R} {\hat \varphi},\tag{27} \\
{\bm B}_{\Psi}=\;& \nabla \times [\nabla \times (\Psi {\hat z})]\\
=\;& \nabla \frac{\partial \Psi}{\partial z} - \nabla^2 \Psi {\hat z}\\
=\;& \frac{\partial^2 \Psi}{\partial R \partial z} {\hat R} + \frac{1}{R} \frac{\partial^2 \Psi}{\partial \varphi \partial z} {\hat \varphi}  - \nabla^2_{\perp} \Psi {\hat z}.\tag {28}
\end{align*}
This representation is the cylindrical version of the Chandrasekhar-Kendall (CK) representation first presented [34] for the field in a spherical domain.  The essential point is to let ${\bm B}_{\Psi}$ account for $B_z$ so that the residual field ${\bm B}_{\Phi}={\bm B}-{\bm B}_{\Psi}$ lies on the $z$ planes, i.e., planes of constant $z$.  As given above the pair $[{\bm B}_{\Psi}, {\bm B}_{\Phi}]$ is not unique.  This field representation can be rendered unique [33,35] for any field ${\bm B}$ in $V$ satisfying boundary conditions (11) and (12) by the following algorithm, referring the reader to the original publications for the details.

First construct the generating function $\Psi$ as a solution of the Neumann boundary value problem (BVP):
$$
\nabla_{\perp}^2 \Psi+B_z(R, \varphi, z, t)=0,\eqno (29)$$
$$
R=R_0, ~~~ {\partial \Psi \over \partial R} = -\frac{F_0}{2 \uppi},\eqno (30)
$$
on any $z$-plane at time $t$ where $\nabla_{\perp}^2$ is Laplacian in that plane.  Boundary condition (30) ensures that ${\bm B}_{\Psi}$ is tangential on $R=R_0$; see eq. (28).  The solution $\Psi$ to this BVP is unique up to a function $\Psi_0(z, t)$, just a constant insofar as the above Neumann BVP is concerned. Since ${\bm B}_{\Psi}$ is invariant under Transformation I: $\Psi \rightarrow \Psi + \Psi_0(z, t)$ for an arbitrary $\Psi_0(z, t)$, it follows that ${\bm B}_{\Psi}$ is uniquely defined.

With no loss of generality, we may set $\Psi_0(z, t) \equiv 0$.  Note that the above Neumann BVP can also be solved on $z=\pm z_0$ where boundary conditions (12) apply.  Denote the BVP solutions by $\Psi_{\pm}(R, \varphi)$ and we may replace boundary conditions (12) by
$$
z=\pm z_0, ~~~ \Psi=\Psi_{\pm}(R, \varphi),\eqno (31)
$$
an equivalent Dirichlet condition.  For contained fields, $\Psi=\Psi_{\pm} \equiv 0$.

By their construction ${\bm B}_{\Psi}$ as well as the residual field ${\bm B}_{\Phi}={\bm B}-{\bm B}_{\Psi}$ are unique, solenoidal and tangential at $R=R_0$.  Since ${\bm B}_{\Psi}$ and ${\bm B}$ have identical $z$-components, ${\bm B}_{\Phi}$ lies in $z$-planes, in the Euler-potential representation (27) with field lines as curves of intersection between flux surfaces of constant $\Phi(R, \varphi, z)$ and the $z$-planes.  With ${\bm B}_{\Phi}$ being tangential at $R=R_0$, its field-lines must close within $R<R_0$ and are unlinked, including the one running along the circular boundary $R=R_0$ expressed by the boundary condition,
$$
R=0, ~~~ {\partial \Phi \over \partial \varphi}=0.\eqno (32)
$$
Transformation II: $\Phi \rightarrow \Phi + \Phi_0(z, t)$ leaves $\Phi(R, \varphi, z, t)$ invariant by eq. (27) for an arbitrary $\Phi_0(z, t)$.  With no loss of generality, we may replace boundary condition (32) with
$$
R=0,\ \Phi=0.\eqno (33)
$$
The construction of the unique pair $[{\bm B}_{\Phi}, {\bm B}_{\Psi}]$ at any instant of time is now complete for ${\bm B}$ as given by eqs. (26)--(28).

The defining property of $[{\bm B}_{\Phi}, {\bm B}_{\Psi}]$ is that the component of ${\bm B}_{\Psi}$ in the $z$-planes is potential; see eq. (28). The circulation of ${\bm B}_{\Psi}$ vanishes around {\it any} closed curve $\gamma$ on a $z$-plane:
$$
\oint_{\gamma} ~ {\bm B}_{\Psi} \cdot {\rm d}{\bm l} = 0,\eqno (34)
$$
and the CK representation may be physically characterized as follows.  The field ${\bm B}_{\Psi}$ accounts for $B_z$ everywhere as a vertical flux passing untwisted through each $z$ plane with zero circulation in that plane.  The circulation of ${\bm B}$ in each $z$-plane is entirely accounted for by the complementary field ${\bm B}_{\Phi}$.

The CK representation is the Eulerian counterpart to the Lagrangian description using the $\tau$ toroidal partition of the fluid in $V$.   A given field ${\bm B}(R, \varphi, z, t)$ in CK representation at $t=t_0$ yields the field ${\bm B}_{\Phi}(R, \varphi, z, t_0)$ to define the partition of the fluid into the unlinked $\tau$ toroids.  Thus we have the identification $\left[{\bm B}_{\tau}, {\bm B}_{\upsilon} \right] _{t=t_0}= \left[{\bm B}_{\Psi}, {\bm B}_{\Phi} \right]_{t=t_0}$.  The Lagrangian description in terms of the $\tau$ toroidal partition then tracts the evolution of $\left[{\bm B}_{\tau}, {\bm B}_{\upsilon} \right]$, each field separately satisfying the induction equation with the common fluid velocity and the two fields together conserving the general Lagrangian helicity ${\mathcal H}$.  In contrast, the CK representation offers an Eulerian description of ${\bm B}$ in terms of an instantaneous decomposition into the linear pair $\left[{\bm B}_{\Psi}, {\bm B}_{\Phi} \right]$ at each time $t$ with no reference to where each fluid parcel is located at that time.   This result shows that the contruction of ${\bm B}={\bm B}_{\tau}+ {\bm B}_{\upsilon}$ in sect. 2.2 is completely general.

\subsubsection{The Euler potentials of the CK fields}

The CK representation is a mathematical proof that any given field is a linear superposition of not more than three solenoidal fields each represented by a pair of global Euler potentials.  This proof follows simply from rewriting eqs. (26)--(28) as:
\begin{align*}
{\bm B} =\;& \nabla \times \left( \Phi {\hat z} \right)  + \frac{\partial^2 \Psi}{\partial R \partial z} {\hat R} + \frac{1}{R} \frac{\partial^2 \Psi}{\partial \varphi \partial z} {\hat \varphi}  - \nabla^2_{\perp} \Psi {\hat z}\\
=\;&  \nabla R \times \nabla \left(\frac{1}{R} \frac{\partial \Psi}{\partial \varphi} \right) + \nabla \varphi \times \nabla  \left( R \frac{\partial \Psi}{\partial R} \right) + \nabla \Phi \times \nabla z,\tag {35}
\end{align*}
displaying the three pairs of Euler potentials explicitly.  Again we note that this field representation
offers two basic means of description,  the first being the Eulerian method in a permanent representation in terms
of $\left[{\bm B}_{\Psi}, {\bm B}_{\Phi} \right]$.  The other is the Lagrangian method that identifies
at some initial time the three fields defined by the pairs of global Euler potentials,
$\left( R, {1 \over R}  {\partial \Psi \over \partial \varphi} \right)$,
$\left( \varphi, R {\partial \Psi \over \partial R} \right)$ and $(\Phi, z)$,
whose evolutions are separately tracked by their Euler potentials moving as fluid surfaces according to advection equations (25).

The presence of multi-valued $\varphi$ as an Euler potential in eq. (35) is a removable feature.  The description of a vector field in terms of scalar functions has the non-trivial advantage that scalar functions have the same value at each point in physical space independent of the coordinate system.   In contrast, the 3 components of a vector field do not preserve their values at a physical point under a transformation between two different coordinate systems.  This advantage is seen in the same field ${\bm B}$ given by eq. (35) taking the following form in Cartesian coordinates $(x, y, z)$:
$$
{\bm B}=-\nabla x \times \nabla {\partial \Psi \over \partial y} + \nabla y \times \nabla {\partial \Psi \over \partial x} + \nabla \Phi \times \nabla z,\eqno (36)
$$
without involving any multi-valued Euler potential.

The result (36) has a practical corollary for numerical MHD.  It is impractical to represent an ergodic field line by approximating ${\bm B}$ as a discrete function of space, such as defined on a fixed computational grid.  The fundamental reason is given in the theory of chaos and integrability in nonlinear dynamics [53] but easy to understand intuitively. Numerically integrating ODEs (10) with a discrete ${\bm B}$ requires extrapolating for the field between the grid points.  Without analytical knowledge of the field being approximated, true ergodicity and the artificial complicated meandering of a computed field line cannot be distinguished, the latter resulting from numerical errors accumulated in the computation.  In other words, ergodicity in a field is information irrevocably lost to a discrete variable.

Each of the three component fields on the right in eq. (36) has global flux surfaces, their ODEs being integrable in the language of chaos theory.  Representing these flux surfaces numerically involves the usual computational and truncation errors, of course.  But, if these geometric surfaces are described with sufficient computational accuracy, information of ${\bm B}$ being ergodic resides in the {\it geometric relationships} among the Euler potentials.  Therefore, ergodicity as a property of ${\bm B}$ is retained as faithfully as the individual Euler potentials are numerically precise.

The computational advantage pointed out here is clear in the fact that the advective equation (25) are the result of an analytical integration of induction equation (5).  This step carries out a pre-computational integration for the field lines instead of performing this integration numerically as a post-computation analysis of a numerical calculation in terms of ${\bm B}$ as a discrete variable.

Having to deal with up to three pairs of Euler potentials instead of ${\bm B}$ seems computationally more intensive.
This concern is mitigated by the manifestly solenoidal form of the fields represented by Euler potentials
and by the advective equation being one of the simplest transport equations to treat numerically.
The linear decomposition (36) has a simple geometric interpretation.  At some initial time, ${\bm B}$
is decomposed into the sum of 3 planar solenoidal fields residing in the respective planes of
constant $x$, $y$, and $z$.  At any subsequent time, ${\bm B}$ is the sum of those 3 fields
deformed by the fluid velocity ${\bm v}$ since the initial time.  Each of
the three fields is untwisted and their mutual entanglements defined by their superposition
describe the twisted topology of ${\bm B}$.  The fundamental computational point here is that
if a high degree of the frozen-in condition is essential for a physical problem [54--57],
the faithful description of this geometric picture, or some equivalent description of this kind, is
what it takes to achieve that essential degree.  The Lagrangian representation eq. (36) has been successfully used in recent studies of CS formation [16,24].

\subsection{Magnetic helicity}

A contained field in $V$ with $B_z(R, \varphi, \pm z_0)=0$ evolves with conservation of the classical total helicity [58,59]
$$
H_{\rm c}\left( {\bm B}; V \right)=\int_V~ {\bm A} \cdot {\bm B} ~{\rm d}V\eqno (37)
$$
under induction equation (5) subject to the electromagnetic conditions.  Although helicity density $h={\bm A} \cdot {\bm B}$ is physically ambiguous because of its dependence on the free gauge $G$ of ${\bm A}$, the helicity $H_{\rm c}$ is gauge independent and physically meaningful.  Under a gauge transformation ${\bm A} \rightarrow {\bm A} + \nabla G$,
$$
H_{\rm c} \rightarrow H_{\rm c} + \int_{\partial V}~ G ~{\bm B} \cdot {\rm d}{\bm S} = H_{\rm c},\eqno (38)
$$
for all gauge function $G$ by virtue of ${\bm B}$ being tangential at domain boundary $\partial V$.
What $H_{\rm c}$ measures is represented by the case of two closed toroidal tubes of magnetic flux,
linked an integer $\pm N$ times about each other, signed for the handedness and ignoring
the internal twist structures of these tubes.  In this case $H_{\rm c}=\pm 2 N f_1f_2$ where
$f_1$ and $f_2$ are the axial fluxes of
the two tubes, i.e., $H_{\rm c}$ may be described as the flux-weighted invariant linkage between two fluxes.  For closed flux tubes, if one goes around the other $N$ times, the converse is true, so $H_{\rm c}$ carries the factor 2 to count both links.

Consider a contained field in $V$ comprising contiguous sub-systems of flux wholly contained within their respective,
outer-most closed flux surfaces.  These containing flux-surfaces could also exist as a continuum of nested closed surfaces, not necessarily simply connected.  In these cases, each of the infinitely-many containing flux-surfaces has a conserved $H_{\rm c}$ under the induction equation [43,45].  The general presence of ergodic field lines in 3D fields presents other possible flux sub-systems.  There
is an infinity of fields in $V$, each comprising a finite number of contiguous sub-systems of flux and each sub-system containing a single, infinitely-long, volume-filling field line.  In a field of this kind, there can be only a finite number of conserved classical total helicities, one for
each of the sub-systems.  From this perspective, these conserved helicities, finite in number, are not
capturing the property of a continuum of conserved fluxes on all fluid surfaces.

For an anchored field in $V$, the classical total helicity $H_{\rm c}$ is not gauge-independent and,
in its place, one may use the total relative helicity $H_{\rm R}({\bm B}, {\bm B}_{\rm ref}; V)$ as a
formal measure of a given field ${\bm B}$ relative to a similar measure of a chosen reference
field ${\bm B}_{\rm ref}$ [38,40,41].  The construction of $H_{\rm R}$ is involved and will not be reproduced here.
Suffice for our purpose here is to recall [38] that the construction renders a difference in helicity
between ${\bm B}$ and ${\bm B}_{\rm ref}$ that is independent of the gauges of the vector potentials of both fields that feature in the formula, the field ${\bm B}_{\rm ref}$ required to have the same boundary-flux distribution as the given field ${\bm B}$.  There is a unique potential field ${\bm B}_{\rm pot}$ in the simply connected $V$ meeting this requirement, which has been generally used as the reference field.  By definition, applying $H_{\rm R}$ to a contained field leads to $H_{\rm R}=H_{\rm c}$ because, in the absence of flux across the boundary, ${\bm B}_{\rm pot} \equiv 0$ for the simply-connected domains considered here.

The CK representation ${\bm B}={\bm B}_{\Phi} + {\bm B}_{\Psi}$ defines a total absolute helicity
\begin{align*}H_{\rm abs}\left( {\bm B}; V \right)=\;&\int_V ~ \left[\left(\nabla \times \Psi {\hat z}\right) \cdot \nabla \times \left( \nabla \times \Psi {\hat z} \right)\right. \\
\;&\left.+ 2 \left(\nabla \times \Psi {\hat z}\right) \cdot \left(\nabla \times \Phi {\hat z}\right)\right]{\rm d}V,\tag {39}
\end{align*}
applicable to the contained and anchored fields on the same conceptual basis.  We describe $H_{\rm abs}$ as absolute to distinguish it from $H_{\rm R}$.  The pair $\left[{\bm B}_{\Phi}, {\bm B}_{\Psi} \right]$ as well as the scalar $H_{\rm abs}$ are invariant under Transformation I: $\Psi \rightarrow \Psi+\Psi_0(z, t)$ and Transformation II: $\Phi \rightarrow \Phi+\Phi_0(z, t)$ for arbitrary $(\Psi_0, \Phi_0)$.  Subject to boundary conditions (30) and  (31) on $\Psi$, boundary condition (32) on $\Phi$, and the electromagnetic conditions (14) on ${\bm v}$, $H_{\rm abs}$ is conserved [33] under induction equation (5).

This development allows us to redefine $H_{\rm R}$ in an absolute sense for the anchored field:
$$
H_{\rm R}\left( {\bm B}, {\bm B}_{\rm pot}; V \right)=H_{\rm abs}\left( {\bm B}; V \right)-H_{\rm abs}\left( {\bm B}_{\rm pot}; V \right),\eqno (40)
$$
giving each of the fields $\left( {\bm B}, {\bm B}_{\rm pot} \right)$ an independent measure of its helicity.  We may regard $H_{\rm abs}$ to be the closing of a conceptual gap remaining in the development from the construction of $H_{\rm c}$ to that of $H_{\rm R}$.  We point out a few interesting implications derived from $H_{\rm abs}$.

An anchored potential field ${\bm B}_{\rm pot}=\nabla P$ satisfying $\nabla^2 P=0$ in the CK representation ${\partial \Psi_P \over \partial z} \equiv P, \Phi \equiv 0$ has the total absolute helicity,
$$
H_{\rm abs}\left( {\bm B}_{\rm pot}; V \right) = \int_V ~ \nabla \times \Psi_P {\hat z} \cdot \nabla \times \left( \nabla \times \Psi_P {\hat z} \right) ~ {\rm d}V.\eqno (41)
$$
If ${\bm B}_{\rm pot}$ is axisymmetric, $H_{\rm abs} \equiv 0$, and we have $H_{\rm abs} \equiv H_{\rm R}$.
If ${\bm B}_{\rm pot}$ is not axisymmetric, for
example, the boundary fluxes  at $z=\pm z_0$ depend on $\varphi$, then generally $H_{\rm abs} \ne 0$.

In such a case, let us fix $B_z=b_{\pm} (R, \varphi)$ at $z=\pm z_0$ and consider a continuous deformation
$V \rightarrow V'$ from a length of $2 z_0$ to a different length, say, $2 z'_0$,
by a uniform compression strictly in the $z$-direction of the fluid and its embedded field,
holding fixed the boundary flux distributions $b_{\pm} (R, \varphi)$. Let ${\bm B}_{\rm pot}$
in $V$ deform into ${\bm B}_{\rm deformed}$ in $V'$. It can then be shown [33] that $H_{\rm abs}$ is conserved,
which is to be expected since the field topology has not changed.
On the other hand, the unique potential field ${\bm B}'_{\rm pot}$ in the new domain $V'$ has a
total absolute helicity $H'_{\rm abs} \ne H_{\rm abs}$, generally.  This result
follows from an interesting property that ${\bm B}_{\rm pot}$ in $V$ and ${\bm B}'_{\rm pot}$ in $V'$ generally are not
topologically identical although both fields share the same boundary flux-distributions.

Take a simple case of $b_{\pm} (R, \varphi)$ being positive and negative definite in $z=\pm z_0$, respectively.  Then each of the field lines has one foot-point at $z=-z_0$ and the other at $z=z_0$ defining a foot-point map.  In such a case, the foot-point map of ${\bm B}_{\rm pot}$ in $V$ is in general distinct [60--62] from the foot-point map of ${\bm B}'_{\rm pot}$ in $V'$.  This property gives an insight into $H_{\rm R}$.  Whereas ${\bm B}_{\rm pot} \rightarrow {\bm B}_{\rm deformed}$ incurs no change in field topology and in total absolute helicity, its $H_{\rm R}$ shows a change in value due not to a change in field topology but, instead, to a change of the reference potential field in the  $V \rightarrow V'$ frozen-in deformation.

Consider the total absolute helicities $H_{\rm abs}$ defined by CK fields in two special domains, the unbounded
space $-z_0<z<z_0$ between infinite planes $z=\pm z_0$ and the finite spherical domain
$r_1<r<r_2$ bounded by a pair of concentric spheres of radii $r_1<r_2$.  In these two
domains [32,33,63--65] $H_{\rm abs} \equiv 0$ for all potential fields, so that $H_{\rm R} \equiv H_{\rm abs}$,
whereas in $V$, $H_{\rm R} \ne H_{\rm abs}$ generally.  A simple topological or physical explanation of this
mathematical result is not available.
That $H_{\rm R} \ne H_{\rm abs}$ generally in the cylindrical domain $V$ is probably representative of most simply connected domains.

All three total helicities $H_{\rm c}$, $H_{\rm R}$ and $H_{\rm abs}$ are Eulerian quantities,
defined according to the state of ${\bm B}$ at a given time with no interest in where each fluid parcel is located at the time.
The CK representation defines a Lagrangian helicity $H_{\rm L}$ in $V$, gauge independent:
$$
H_{\rm L} \rightarrow H_{\rm L} + \int_{\partial V}~ G {\bm B}_{\tau} \cdot {\rm d}{\bm S} = H_{\rm L},\eqno (42)
$$
by virtue of ${\bm B}_{\tau}$ being tangential at boundary $\partial V$.  Note that all four helicities
$(H_{\rm c}, H_{\rm R}, H_{\rm abs}, H_{\rm L})$ have the
same physical dimension of the square of magnetic flux.  This helicity $H_{\rm L}$ is well defined irrespective of whether the given field ${\bm B}$ is contained or anchored.  We note again that $H_{\rm L}$ is complicated to compute, requiring knowledge of the location of every fluid parcel identified by a specific $\tau$ partition of the fluid.  If this Lagrangian information is available, we can calculate a two-parameter set of conserved fluxes $\left[ f(\tau), F(\tau, \upsilon) \right]$, in terms of which a continuum of general helicity ${\mathcal H}$ can be stipulated, of which $H_{\rm L}$ is a particular case.

The existence of ${\mathcal H}$ suggests that in addition to these simple Eulerian helicities,
$H_{\rm c}$, $H_{\rm R}$, $H_{\rm abs}$, $H_{\rm L}$, there exists a corresponding
continuum of general Eulerian helicity ${\mathcal H}^*$ conserved for the whole
volume $V$ under perfect conductivity.  Worthy of note is that ${\mathcal H}$ by its
definition is a summation over all the $\tau$ toroids but, with the exception of
$H_{\rm L}$, ${\mathcal H}$ is generally not a simple integral over the domain $V$.
Although the formidable mathematical complexities of ${\mathcal H}$ and ${\mathcal H}^*$ are
interesting in their own rights, with useful applications in certain problems, our review
in this section reminds us that they are equivalent properties derived from the simple
Lagrangian statement of flux conserved on {\it all} fluid surfaces.
This understanding is built upon translating the induction equation directly into a
physically complete description of ${\bm B}$ in terms of a superposition of up
to three topologically simple solenoidal fields.  We next present a physically contrasting system where such
a description is also useful, before turning to treat near-ideal MHD.

\subsection{Ideal vortex dynamics}

Consider an incompressible inviscid fluid in the absence of the magnetic field.  Momentum and mass-conservation equations (1) and (3) reduce to the Euler equations:
$$
\frac{\partial {\bm v}}{\partial t} + \left({\bm v} \cdot \nabla \right) {\bm v} = -\nabla p, \eqno (43)$$
$$
\nabla \cdot {\bm v}=0,\eqno (44)
$$
\noindent
where the uniform density of the fluid has been set to unity. Rewrite eq. (43) as:
$$
\frac{\partial {\bm v}}{\partial t} + (\nabla \times {\bm v}) \times {\bm v} = -\nabla \left( p + \frac{1}{2} v^2 \right),\eqno (45)$$
from which the pressure can be eliminated by taking the curl across:
$$\frac{\partial {\bm w}}{\partial t} = \nabla \times ({\bm v} \times {\bm w}),\eqno (46)
$$
introducing the vorticity
$${\bm w}= \nabla \times {\bm v}.\eqno (47)
$$
For each solution ${\bm v}$ of eq. (46), the pressure is determined at each instant in time by the Poisson equation:
$$
\nabla^2 p = - \nabla \cdot \left[ \left({\bm v} \cdot \nabla \right) {\bm v} \right],\eqno (48)
$$
given by the divergence of eq. (43).  Consider the following alternative to the above standard treatment.

Vorticity equation (46) has the form of induction equation (5) identifying ${\bm B}$ with ${\bm w}$, as is well known.
It follows from our development in this paper that the incompressible vorticity ${\bm w}$ has the CK representation:
$$
{\bm w}=\sum_{i=1}^3 ~w_i(\xi_i, \zeta_i) \nabla \xi_i \times \nabla \zeta_i,\eqno (49)$$
$$\left[\xi_1, \xi_2, \xi_3 \right]= \left[ x, y, z \right],\eqno (50)$$
$$\left[\zeta_1, \zeta_2, \zeta_3 \right]= \left[-\frac{\partial \Psi}{\partial y}, \frac{\partial \Psi}{\partial x}, -\Phi \right],\eqno (51)
$$
at any given instant of time $t_0$, in a simple application of eq. (36).  If we pick one such representation and keep it for all time $t>t_0$, we have a Lagrangian description of ${\bm w}$ as a linear superposition of 3 topologically simple solenoidal fields.   Simplicity is meant that each field is expressible in terms of a pair of globally defined Euler potentials.

Define
$$
w_i(\xi_i, \zeta_i) = \frac{\partial \omega_i(\xi_i, \zeta_i)}{\partial \xi_i},\eqno (52)
$$
in order to obtain three equivalent pairs of Euler potentials,
$$
{\bm w}=\sum_{i=1}^3 ~\nabla \omega_i (\xi_i, \zeta_i) \times \nabla \zeta_i= \frac{1}{2}\sum_{i=1}^3 ~ \nabla \times \left( \omega_i \nabla \zeta_i - \zeta_i \nabla \omega_i \right).\eqno (53)
$$
By definition eq. (47), the velocity is explicitly given by
$$
{\bm v}=\nabla W + \frac{1}{2}\sum_{i=1}^3 ~\left[ \omega_i \nabla \zeta_i - \zeta_i \nabla \omega_i \right],\eqno (54)
$$
introducing an arbitrary potential $W$ which must satisfy the Poisson equation,
$$
\nabla^2 W =- \frac{1}{2}  \sum_{i=1}^3 ~\nabla \cdot \left[ \omega_i \nabla \zeta_i - \zeta_i \nabla \omega_i \right],\eqno (55)
$$
under incompressibility.

Euler equation (43) then yields the pressure
$$
p=  p_0 + \frac{\partial W}{\partial t} - \frac{1}{2}  \sum_{i=1}^3 ~\left[\omega_i \frac{\partial \zeta_i}{\partial t}- \zeta_i \frac{\partial \omega_i}{\partial t} \right]- \frac{1}{2} v^2,\eqno (56)
$$
$p_0$ being an integration constant, and reduces to the advection equation (25) rewritten for the six Euler potentials $\left[\omega_i, \zeta_i\right], ~i=1, 2,3$ moving with the common fluid velocity ${\bm v}$ given by eq. (54).  These advection equations together with Poisson equation (55) constitute a complete set of 7 governing PDEs for 7 scalar unknowns, subject to the rigid boundary conditions on ${\bm v}$.

As in the corresponding MHD problem, the advection equations for the six Euler potentials constitute an analytical integration of the vorticity equation (46).  This integration allows a high degree of frozen-in vorticity to be described numerically, as high a degree as the computed Euler potentials are numerically accurate.   This level of accuracy is essential in the simulations of formation of singularities in ideal vortex dynamics in parallel to CS formation in ideal MHD.

On a historical note, Clebsch (1857) was the first to represent incompressible inviscid vorticity in terms of a pair of Euler potentials [66] but little development ensued from this important work because the single-pair representation is in general not global.  This neutral fluid system is an instructive contrast to ideal MHD in the correspondence between $({\bm w}, {\bm v}, W)$ and $({\bm B}, {\bm A}, G)$.  Whereas $({\bm A}, G)$ are physically ambiguous because $G$ is an arbitrary free gauge, the corresponding pair $({\bm v}, W)$ are physically unique quantities.  If limited to a single pair of Euler potentials to represent ${\bm B}$, this representation is generally local because the Euler-potential pair generally are multivalued scalar functions of space, requiring distinct Euler-potential pairs to describe ${\bm B}$ in different spatial regions of applicability.  The issue of the corresponding multivalued nature of vector-potential ${\bm A}$ can be bypassed by not using it.  For example, the CK representation in sect. 2.3.2 expresses ${\bm B}$ in terms of $({\bm B}_{\Psi}, {\bm B}_{\Phi})$ as uniquely defined single-valued vector functions of space.  In the case of the neutral fluid the velocity ${\bm v}$, as the vector potential of ${\bm w}$, and the potential $W$, as the equivalent of the free magnetic gauge $G$, both have direct physical meanings.  In both cases, we have a completely general, global, linear decomposition of the field, ${\bm B}$ or ${\bm w}$, into three simple solenoidal fields.

\section{Breaking of field topology in weakly resistive fluids}

The motion of a perfect fluid in 3D space may be visualized locally as the interactions among contiguous magnetic flux tubes.
No exchange of fluid among the flux tubes is allowed.  The interaction among three (non-parallel) tubes in the absence of artificially imposed geometric symmetries can produce a TD or CS; Figure 1.  Two tubes pressing into a third tube between them can readily push it clear out of the way to meet partially at a contact flux-surface.  The nonlinear dynamics of a particular situation determines which three tubes in the continuum field should behave in this manner. Magnetic neutral point defined by ${\bm B}=0$ have no special significance [67].  What is essential in this generally 3D process is the interaction of three or more locally distinct flux systems.

\end{multicols}

\begin{figure}[H]
\centering
\includegraphics[scale=0.95]{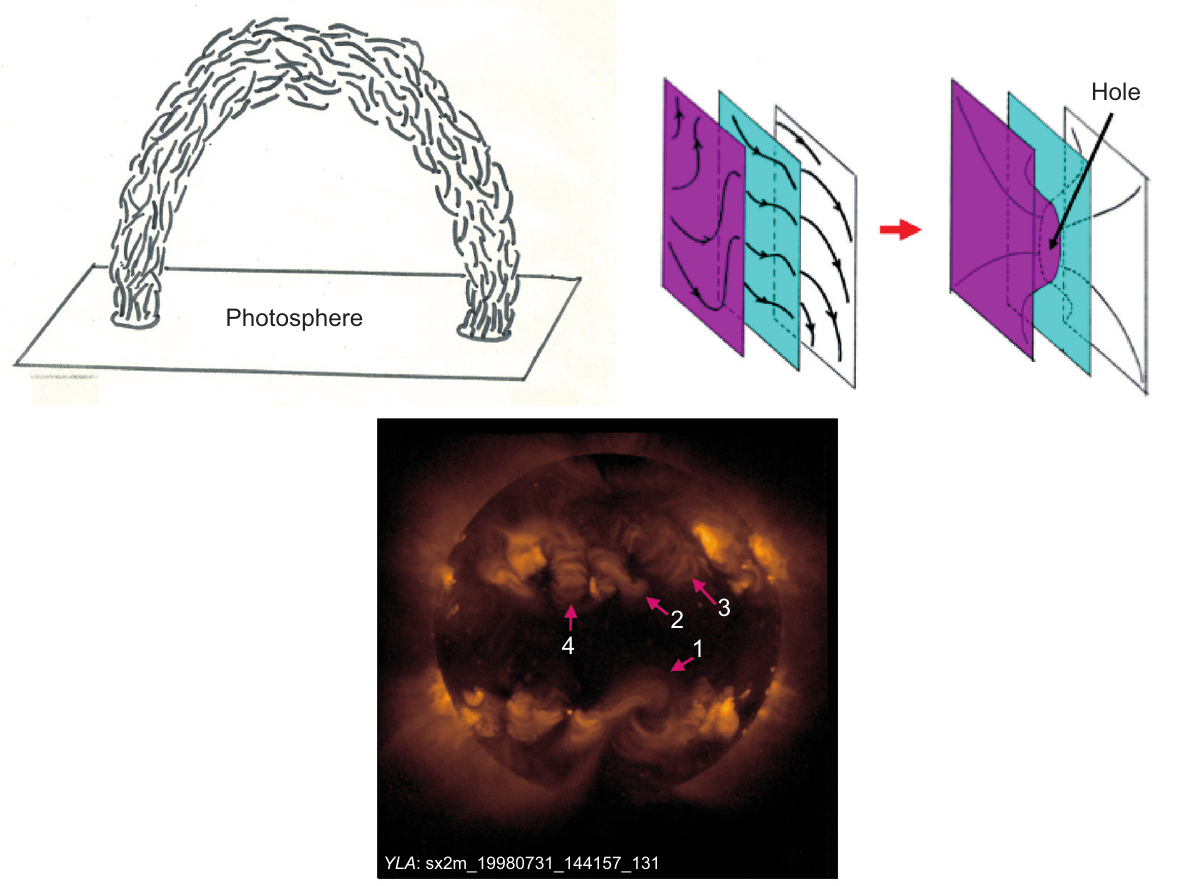}
\caption{Relating the Parker Magnetostatic Theorem (top) to the (2--3)$\times 10^6$ K corona
(bottom) observed in X-ray emission [82].  Sketch of a coronal bipolar flux tube
(top left) with footprints anchored to the photosphere, the visible solar surface,
with the tangled state of the field lines suggesting that preserving such a complex field topology
would, during relaxation, result in TD formation illustrated by the inward collapse of the colored
flux surfaces (top right), creating the hole in the central blue-colored flux surface,
as described in the text.  Structures numbered 1 and 2 in the X-ray corona display forward and reflected
 S plasma morphologies discussed in the text.  The sketch of the bipolar flux tube
is taken with permission from Parker E N, Rapid Reconnection, In: Gonzalez W,
ed. Parker Reconnection Workshop 2014. New York: Springer,  2015.  The image
of the corona is publicly available from the {\it Yohkoh} Mission of Institute of
Space and Astronautical Science (ISAS), Japan, made with the Soft X-ray Telescope
prepared by the Lockheed-Martin Solar and Astrophysics Laboratory, the National
Astronomical Observatory of Japan, and the University of Tokyo with the support
of US National Aeronautics and Space Administration (NASA) and ISAS.
} %ͼ��
\label{fig:example2}
\end{figure}

\textwidth=178truemm \textheight=236truemm%%%%%%�°�ʽҪ����

%%%%%%%%%%%%%%%%%%%%%%%%%%%%%%%%%%%%%%%%%%%%%%%%%%%%%%%%%%%%
\wuhao\vspace*{2mm}

\begin{multicols}{2}

%%%%%%%%%%%%%%%%%%%%%%%%%%%%%%%%%%%%%%%%%%%%%%%%%%%%%%%%%%%%
%% Text of article.
%%%%%%%%%%%%%%%%%%%%%%%%%%%%%%%%%%%%%%%%%%%%%%%%%%%%%%%%%%%%
%    Section headings
\renewcommand{\baselinestretch}{1.08} \baselineskip 12.2pt\parindent=10.8pt

\renewcommand{\thefootnote}

An extremely highly conducting fluid behaves similarly except that the monotonic steepening of a field at a contact surface proceeds to a point of dissipation at the small nonzero resistivity.  Resistive magnetic reconnection sets in and field topology is no longer preserved.  Thus the highly conducting fluid spontaneously creates dynamical situations in which resistive dissipation is inevitable.\vspace{-0.5mm}

The perfect fluid conductor as a singular limit of extremely highly conducting fluids derives from the Parker Magnetostatic Theorem reviewed next.  This theorem states that the static equilibrium of a 3D field imposed with an arbitrarily prescribed topology $T$ generally must contain TDs as an intrinsic component of the equilibrium.  The discrete (sheet) currents flowing in the TDs and the continuous part of the current density, subject to boundary conditions, together determine the distribution of the equilibrium ${\bm B}$ by Ampere's law.  For this equilibrium field to possess topology $T$, currents in the form of TDs must be present for most prescribed $T$.  The complete development of this fundamental theorem is presented in the monograph of Parker [8] and recent review articles [9,10,17,22]. In this section we concentrate on basic concepts.\vspace{-0.5mm}

The inevitability of TD formation for a fixed $T$ extends beyond static consideration.  Even in a dynamical state, TDs or CSs may form, perhaps in finite time [54,55], in a locality without the field attaining equilibrium everywhere.  The dynamical forces of the rich time-dependent processes of MHD may act both ways depending on circumstance, to aid or to frustrate the formation of a TD.  Therefore the Parker theorem is saying that, even if the complexity of these time-dependent MHD processes are removed, TDs are generally still inevitable if a field is to just attain equilibrium everywhere in a domain with its given $T$.

Time-dependent 3D MHD numerical computation is a powerful practical means of demonstrating spontaneous formation of TDs.  Differential calculus is the basis of most numerical computations.   To successfully demonstrate a TD developing under the frozen-in condition to the verge of losing differentiability requires mature understanding [57] of the computational demands of such a numerical undertaking [16,20,24], as pointed out in sect. 2.3.3.

\subsection{Parker Magnetostatic Theorem}

Set $\eta=0$ for a perfect fluid conductor in a cylindrical domain $V$ and let the fluid be viscous, with $(\nu_1, \nu_2) \ne 0$ in momentum equation (1).  For our purpose consider a cold gas, setting pressure $p=0$.  The closed system of equations, (1), (3) and (5), determines $({\bm B}, {\bm v}, \rho)$ subject to boundary conditions (11) and (12) and rigid frictional boundary condition (15).  Contained and anchored fields are treated on the same basis here.  The problem being addressed is the end state of the viscous relaxation we have set up.

The field exerts a Lorentz force on the fluid as kinetic energy is continuously dissipated by viscosity.
Field topology $T$ defined by an
initial field at time $t_0$ is invariant in time. By artificially not feeding the lost kinetic energy back into the fluid, the total energy $E$ of the system
\begin{align*}
E =\;& E_{\rm M}+E_{\rm K}, \\
E_{\rm M}=\;&\frac{1}{8 \uppi} \int_V~ B^2 ~ {\rm d}V,\\
E_{\rm K}=\;&\frac{1}{2} \int_V~ \rho v^2 ~ {\rm d}V,\tag {57}
\end{align*}
decreases monotonically in time starting from its bounded value $E=E_0$  at time $t_0$.  The invariant $T$ implies that the field cannot be completely removed whereas kinetic energy is being removed irretrievably at the expense of field energy.  Therefore, $E_{\rm K} \rightarrow 0$ as $E_{\rm M}$ tends to a minimum, defining the end state in which both ${\bm v}$ and the Lorentz force vanish, described by the force-free equations
\begin{align*}
\left( \nabla \times {\bm B} \right) \times {\bm B} =\;& 0, \tag {58}\\
\nabla \cdot {\bm B} =\;& 0,\tag {59}
\end{align*}
subject to boundary conditions (11) and (12) and to a fixed topology $T$.

The following is a mathematically useful restatement of the Parker theorem [22].  Consider the space ${\mathcal B}$ of all {\it continuous} fields in $V$, not necessarily in a force-free state, satisfying the given boundary conditions (11) and (12).  Define the continuum set ${\mathcal T}$ of all the field topologies $T$ realized in these fields.  By definition ${\mathcal B}=\cup_T ~ {\mathcal B}_T$, the union of all the disjoint subspaces ${\mathcal B}_T$ of fields, each subspace containing fields of a common topology $T$.  It is formidable to stipulate $T$ in explicit mathematical terms [68--74], as is clear from sect. 2.   However, $T$ is conceptually unambiguous if defined by its realization in a specific field in $V$ from which all members of the subspace ${\mathcal B}_T$ can be generated by all the continuous velocities under the ideal induction equation subject to boundary-condition (15).

Next consider the subspace ${\mathcal B}_{\mathrm {ff}}$ containing all the {\it continuous} force-free fields with no TDs, satisfying boundary conditions (11) and (12).  Denote by ${\mathcal T}_{\mathrm {ff}}$ the continuum of field topologies realized in the force-free fields in ${\mathcal B}_{\mathrm {ff}}$.  By definition ${\mathcal B}_{\mathrm {ff}} \subseteq {\mathcal B}$ and ${\mathcal T}_{\mathrm {ff}} \subseteq {\mathcal T}$.  The Parker theorem may then be stated that for most 3D fields in $V$, ${\mathcal B}_{\mathrm {ff}}$ and ${\mathcal T}_{\mathrm {ff}}$ are subsets of measure zero against their respective mother sets ${\mathcal B}$ and ${\mathcal T}$.

A random pick of topology $T_1 \in {\mathcal T}$ has an unlikelihood of belonging to ${\mathcal T}_{\mathrm {ff}}$.
Consider then $T_1 \notin {\mathcal T}_{\mathrm {ff}}$.  A relaxation starting from an initial state
in ${\mathcal B}_{T_1}$ must evolve with a monotonically decreasing $E$ into
an end-state force-free field ${\bm B}_{t \rightarrow \infty}(T_1)$ containing TDs.
i.e., the limit point ${\bm B}_{t \rightarrow \infty}(T_1) \notin {\mathcal B}_{T_1}$.
In mathematical analysis, a function space is called Cauchy complete if all
convergent infinite sequences of points in the space converge to end-points
belonging to the space.  Most functional spaces are not Cauchy complete [75].
The open interval on a real axis is an one-dimensional example of a space not Cauchy complete, and more
complicated examples are not difficult to construct [22,76].  Proving the property ${\bm B}_{t \rightarrow \infty}(T_1) \notin {\mathcal B}_{T_1}$ is not trivial in any particular problem whereas the property of a function space not Cauchy complete is common in variational calculus [77,78].

The essential point here is that the existence of a minimum of the integral $E_{\rm M}$ over the
function space ${\mathcal B}_{T_1}$ does not imply that the minimum value must be
realized in that space.  Viscous relaxation drives a field preserving topology $T_1$
unavoidably to a minimum in $E_{\rm M}$.  Sincere $T_1 \notin {\mathcal T}_{\mathrm {ff}}$,
the relaxation then takes a path of monotonically decreasing $E_{\rm M}$ that is located
in ${\mathcal B}_{T_1}$ except for the path's end-point
${\bm B}_{t \rightarrow \infty}(T_1) \notin {\mathcal B}_{T_1}$.
Intuitively one may think of the analogy of an open interval on the real line.
It bears emphasis that TDs and CSs are admissible in a perfect conductor, so fields of topology $T_1$ containing TDs are physically meaningful.   This end-state is a weak solution of the force-free equations, the TDs satisfying these PDEs in the integral sense [78].  The hydrodynamic shock given by the Rankine-Hugoniot conditions in compressible hydrodynamics is a familiar example of a weak solution.

Fundamental to the theorem is the three-tube interaction.  The coming together of two tubes, squeezing the fluid and its frozen-in field out from between them, results in a hole punched in the contact flux surface where the two flux tubes meet; see Figure 1.  This process is neatly illustrated by the optical analogy constructed by Parker [76,79,80].  Introduce the scalar function $\alpha$ relating the current density to its parallel field and rewrite the force-free equations as:
\begin{align*}
\nabla \times {\bm B} =\;& \alpha {\bm B},\tag {60}\\
{\bm B} \cdot \nabla \alpha =\;& 0.\tag {61}
\end{align*}
Therefore every magnetic flux surface is also a current-density flux surface in a force-free field.
The absence of a curl of the field perpendicular to any flux surface implies that the field is potential
in each flux surface, the field lines distributed exactly like light rays
governed by a corresponding Fermat's principle.  The refractive index is proportional to $|{\bm B}|$, refracting field lines away from regions of strong field strengths.  Thus the field lines of a force-free field on a flux surface may be completely excluded a region of surficiently strong $|{\bm B}|$.  These exclusion regions are the holes in the flux surface in 3D space where field lines external to the particular flux surface stream in from either side to meet tangentially along a contact TDs.

Another physical approach to the Parker theorem investigates an infinitesimally small neighborhood of a given force-free field ${\bm B}$ of topology $T$ in the function space ${\mathcal B}_{\mathrm {ff}}$ of continuous force-free fields.   This neighborhood contains fields of  topologies infinitesimally different from $T$ by some measure.   Using nonlinear perturbational analysis [6,18] it can be shown that, in general, the topologies in this neighborhood in ${\mathcal B}_{\mathrm {ff}}$ is a subset of measure zero against the topologies of the fields in the bigger neighborhood of the given field that extends from within ${\mathcal B}_{\mathrm {ff}}$ into the mother space ${\mathcal B}$.

The stability with which the coronae of the Sun and billions of solar-like stars in our Galaxy maintain
their observed million-degree temperatures finds an attractive MHD explanation in the Parker theorem [8,29].  In the case of the Sun, the field is of the order of 10 G in the low corona, commonly in the form of a bipolar magnetic loop with a pair of foot-prints on the solar surface called the photosphere, sketched in Figure 1.  At its tenuous proton density of $10^{8-9}$ cm$^{-3}$, the ratio $\beta$ of fluid to magnetic pressures is well less than unity.  The field is a ready source of energy for heating and maintaining the corona's temperature, provided, of course, the electric current in the highly conducting corona can be dissipated at its low resistivity.  The free electrons of the fully ionized corona conduct heat along the magnetic field but not across the field.  Therefore, the magnetic flux surfaces are essentially thermal insulators.  The two features of (1) heating despite low resistivity and (2) ubiquitous supply of heat to flux tubes thermally insulated one from another, are explained naturally by the Parker theory of coronal heating [8,81].  The significance for the latter feature is that all flux
tubes have equal likelihood of being heated because any flux surface may have a hole punched into it to form a CS in the general 3D dynamical situation.  The dissipation of spontaneous CSs in a turbulent sea of reconnections breaks field topologies on the small scales to maintain the 2 to 3 million degree temperature of the corona.  As shown by the X-ray image in Figure 1 taken from a recent publication [82], the corona at high activity is characterized with long-lived macroscopic structures of enhanced heating and density.  The numbered structures identify two types: Nos. 1 and 2 display large-scale plasma loops that are conspicuously twisted [83--85] whereas Nos. 3 and 4 display relatively untwisted plasma loops.  The relationship between such large-scale organized magnetic structures and the ubiquitous reconnections on the small scales is the subject of our next subsection.

\subsection{Taylor hypothesis and its generalization}

With each chaotic magnetic reconnection brought about by the dissipation of a CS, the reconnected field is as likely to be topologically compelled to form CSs as the pre-reconnection field that produced the newly dissipated CS.  So CSs would form and dissipate in a perennial turbulent state.  In a recent 3D numerical simulation [16] of the Parker theorem, the viscous evolution to the first formations of CSs involves a laminar continuous velocity but, upon the artificial dissipation of the CSs forming, resulting from numerical truncations, the computed field and flow rapidly develop into a turbulent state; see Figure 18 in this study.   Worthy of note is that the intensity of a CS forming depends on the free magnetic energy available, so the inevitability of a CS forming in a given situation is separate from how energetic the consequent reconnection is. It is conceivable that as the free energy drains away and there is no external source of energy, CSs may still be expected to form persistently but at monotonically weaker intensities [62].

To fix ideas, consider an anchored field in the cylindrical system $V$, taking both the rigid boundary and fluid to be perfectly conducting for the present.  The unique potential field ${\bm B}_{\rm pot}$ has the lowest total energy $E_{\rm pot}$ among all the fields in ${\mathcal B}$ admissible in $V$.  A given field ${\bm B}$ of topology $T$ and total energy $E_{\rm M}$ in $V$ by the Parker Magnetostatic Theorem defines an absolutely-minimum energy state ${\bm B}_{t \rightarrow \infty}(T)$ with total energy $E_{\infty}(T) \ge E_{\rm pot}$, using the notations in sect. 3.1.  Therefore the field ${\bm B}$ of topology $T$ has the free energy $\Delta E_{\rm M}(T)=E_{\rm M}-E_{\infty}(T)$.

Now keep the boundary of $V$ as a rigid perfect conductor but let the fluid in $V$ be weakly resistive so that
CSs form and dissipate resistively during a dynamical evolution of the field ${\bm B}$.
With breakage of field topology, both topology $T$ and free energy $\Delta E_{\rm M}(T)$ vary with time as
${\bm B}$ evolves along a path that lies in the wide open space of ${\mathcal B}$ instead of being
confined within a subspace of fields of a fixed common topology.
Nevertheless the highly conductive fluid is distinct from a resistive static medium.
The high-conductivity driving the formation of CSs for resistive dissipation also sets
macroscopic limits on the amount of free energy that can be removed via CS dissipation.
Strong Faraday induction readily produces strong fresh electric-currents in response to all,
ideal and resistive, changes in the field.  Whereas a rigorously ideal fluid must conserve a
continuum of general helicity ${\mathcal H}$, none is conserved in a static highly-resistive medium.
The intuition follows that perhaps some suitably defined set of general helicities may be
approximately conserved in the limit of $\eta \rightarrow 0$, the essence of the Taylor hypothesis [42,44,45].

The axial flux $F_0$ remains a constant in time since the rigid boundary is perfectly conducting.
Note that $F_0$ is a special case of the general helicity ${\mathcal H}$.
By definition $F_0= f(\tau_1)$ where $\tau_1$ denotes any one of the infinitesimally thin toroids of fluid
that runs along the boundary $R=R_0$. In principle most other forms of ${\mathcal H}$ are not conserved.  To capture the powerful inductive effects of a near-perfect conductor, the Taylor hypothesis postulates that $H_{\rm abs}$ is {\it approximately} conserved in addition to $F_0$.  This postulate constrains the viscous relaxation of ${\bm B}$ in ${\mathcal B}$ to be along a path of constant $F_0$ and monotonically decreasing total energy $E_{\rm M}$ with $H_{\rm abs}=H_0$, a prescribed constant.  Field topology is changing along this path.  Unless the constant $H_0$ is compatible with a potential field, a certain amount of current is always present in the evolving field under the condition $H_{\rm abs}=H_0$ and, in this case, the end state is a non-potential force-free field.  Since $\eta \ne 0$, this end-state contains no TDs.

A recent study [35] treated the variational problem of $\delta E_{\rm M}=0$ subject to a
constant $F_0$ and $H_{\rm abs}=H_0$ over the space ${\mathcal B}$ to show that the minimum-energy
Taylor state in $V$ is governed by the linear force-free equation (60) for $\alpha=\alpha_0$, a constant,
subject to boundary conditions (11) and (12), the solenoidal condition (61) trivially satisfied in this case.
Determining the constant $\alpha_0$ for a 3D system is a mathematically formidable task,
the 3D form of the problem unavoidable if, for example, the prescribed boundary fluxes $b_{\pm}(R, \varphi)$
of the anchored field are $\varphi$-dependent.  The boundary value problem for the linear force-free
field is governed by a linear scalar integro-PDE [35], a novel result contrary to a long-held but erroneous
expectation that all linear force-free fields in $V$ are governed by the Helmholtz PDE [86--89].  For fixed $H_{\rm abs}=H_0$ and $F_0$, the force-free integro-PDE subject to boundary conditions determines a spectrum of admissible $\alpha_0$, analogous to the classical eigenvalue problem.  Each admissible $\alpha_0$ describes an extremum $\delta E_{\rm M}=0$ force-free field.  The absolutely-minimum $E_{\rm M}$ Taylor state is then to be found among these extremum states.

The Taylor hypothesis assumes ubiquitous magnetic reconnections.  In the presence of an extremely high electrical conductivity, resistive dissipation takes place in the exceedingly small space-time volumes of individual CSs.  So resistive dissipation of both $H_{\rm abs}$ and total energy $E_{\rm M}$ is a higher order effect in this sense. Ideal motions between events of CS dissipation leave $H_{\rm abs}$ unchanged whereas total energy $E_{\rm M}$ can change by ideal work, of either sign, done by the field.   A change in topology $T$ by reconnection changes the free energy $\Delta E(T)$ available for the work done, associated with a negligible change in $H_{\rm abs}$.  In a compressible turbulence, MHD shocks provide a ready means of dissipating kinetic and magnetic energies.  This implies an $H_{\rm abs}=H_0$, constant-$F_0$ evolution progresses with a statistically monotonically decreasing $E_{\rm M}$ in the space ${\mathcal B}$.

We digress here to recall that the Taylor hypothesis was original formulated for the contained field in a
simply connected domain like $V$, for which the classical total $H_{\rm c}$ is meaningful and
held constant under the hypothesis.  For the anchored field, $H_{\rm c}$ is not a valid
measure but the Taylor hypothesis may be recovered by replacing $H_{\rm c}$ with the
relative total helicity $H_{\rm R}$, the latter a measure relative to its associated unique potential
field ${\bm B}_{\rm pot}$.  With the theoretical discovery [33] of $H_{\rm abs}$, $H_{\rm R}$ for an
anchored field may be given an interpretation as the difference (40) between the total absolute
helicities of the given field and its unique potential field, both independently evaluated.
Here we run into an interesting issue [17,33,35,90,91], pointed out in sect. 2.4 that the total absolute helicity of a 3D potential field is not necessarily zero.  We need to understand these novel properties in elementary terms and re-examine the problems of defining the Taylor minimum-energy state in terms of $H_{\rm R}$ and $H_{\rm abs}$.

The Taylor hypothesis remains conceptually meaningful independent of such technical details as mentioned above.  In general physical
terms, this hypothesis makes two points about the highly inductive fluid with a weak resistivity.  By the Parker theorem, this fluid is efficiently dissipative via the spontaneous CSs.  On the other hand, the strong induction of Faraday in the limit of $\eta \rightarrow 0$ does not allow all the free magnetic energy to be discharged via CSs.  Here we have a mechanism of storing magnetic energy in the form of field-aligned currents in the solar corona [46].

The Sun's global magnetic field undergoes global polarity reversal every eleven years, marked by the
appearance of a new generation of sunspots at the beginning of each cycle [13,14,47--50,82].   The violence of solar activities in the forms of flares and coronal mass ejections are the consequences of two global fields of opposite polarities having to mix to reach a new equilibrium in the electrically highly-conducting corona [13,14,92--100].  Within the first 5 years of a new cycle, the old coronal global field of a particular polarity is reconfigured with the emerged new field into a similar dipolar form of the opposite global polarity.  Not only are large amounts of magnetic energy liberated in episodes during this large-scale evolution of the corona, the formation of long-lived
structures that build up and store those amounts of energy is an integral part of the global phenomenon.  This rich phenomenology is outside the scope of the review.  The following two points suffice for the purpose of this review.  The significance of the long-lived structures Nos. 1 and 2 in Figure 1 is that their plasma loops indicate twisted, non-potential magnetic fields [83--85].  Figure 2 presents another class of long-lived large-scale coronal structures, the quiescent prominences [101--113], to give specificity to our preceding general remarks on long-lived coronal structures.

To generalize the Taylor hypothesis, one possibility is to find a basis for imposing additional helicities,
selected from the continuum of general helicity ${\mathcal H}$, to define the end state of a turbulent
MHD relaxation, extending an earlier study of the contained field [43].  How is the weak breakdown of
the flux-conservation law to be formulated from first principles in the limit of $\eta \rightarrow 0$?
This is a problem of the coupling between the resistive induction equation and the
dynamics of CS formation and dissipation.  Numerical MHD modeling is a general practical means of exploring these questions, but motivation and guidance with insightful analytical ideas seem essential.

Finally, the Taylor hypothesis also needs to be extended to fields in an open
corona [82,114--117].  There is no intrinsic upper limit to the helicity $H_{\rm abs}$ of a
field confined by rigid walls into a finite domain. In contrast, a force-free field anchored
to the base of an open atmosphere has a stringent upper bound to the free energy
it can store [13,14,118].
Related to this energy bound is a conjecture that in the absence of rigid walls,
a large-scale field low in the corona cannot self-confine when its accumulated helicity
is excessive by some MHD measure[13,14,98,119--125].
The low-coronal magnetic structures not only dump significant energies as flares whenever a significantly-lower energy state becomes available for a helicity-conserving transition in the course of evolution.  A significant part of an entire structure may also lose self-confinement and be ejected into the solar wind [126] carrying along its excessively accumulated helicity [13,14,95,98,99,127--132].
\end{multicols}

\begin{figure}[H]
\centering
\includegraphics[scale=0.95]{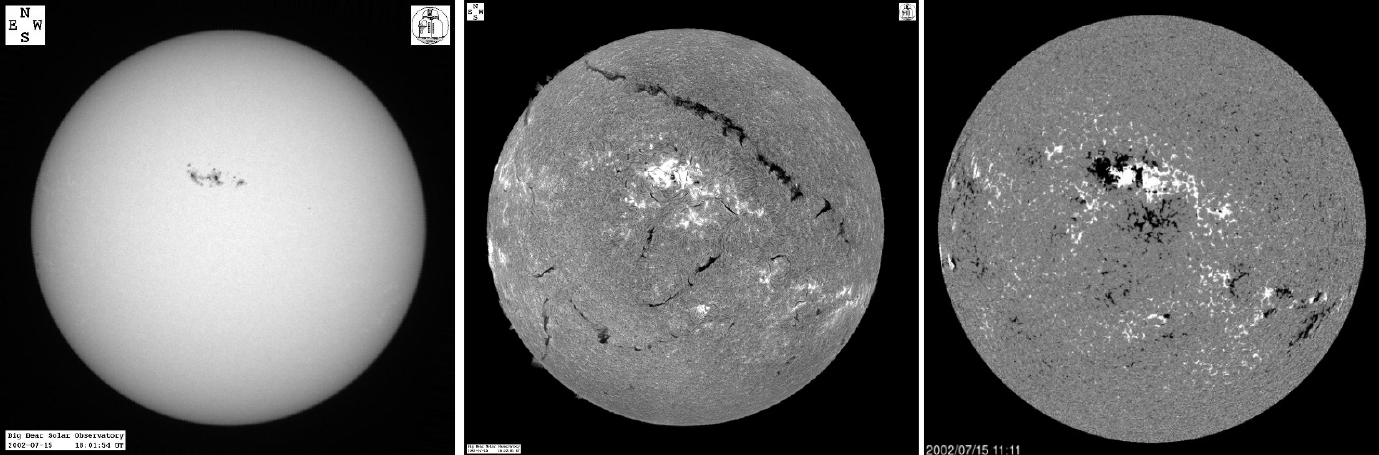}
\caption{The global state of the Sun on July 15, 2002.
The white-light solar surface, the photosphere (left), has a cluster of dark sunspots lined
up along a northern low latitude, north at top of subfigure.
The line-of-sight component of the Sun's photospheric magnetic field (right) indicates the strong
$\approx 10^3$ G fields associated with the sunspots, white and black for positive and negative polarities,
respectively.  Elsewhere globally the field is weak of the order of 10 G, in the form of peppery
small-scale bipolar sources.  In $H_{\alpha}$ emission (center) from a thin layer, the chromosphere,
above the photosphere, the general mottled appearance is due to local magnetic structures,
with bright heated regions over the sunspot cluster as well as quiescent, lengthy condensations,
dark in $H_{\alpha}$ absorption, called prominences.  The conspicuously lengthy
prominence in the northern hemisphere, extending horizontally more than a solar radius
across the solar disk, is two orders of magnitude denser and cooler than the surrounding
tenuous hot corona over weak-field regions away from the sunspots.  Current interpretation
suggests that such a prominence is embedded in a horizontal rope of twisted field [109--112]
that has self-organized out of the remnant, still-twisted fields of sunspots that have decayed.
The left and central subfigures are publicly available from the Big Bear Solar Observatory,
New Jersey Institute of Technology, USA.  The right subfigure is publicly available from the
{\it Solar and Heliospheric Observatory} Mission of the European Space Agency and NASA,
made with the Michelson Doppler Imager [113].
} %ͼ��
\label{fig:example2}
\end{figure}

\textwidth=178truemm \textheight=236truemm%%%%%%�°�ʽҪ����

%%%%%%%%%%%%%%%%%%%%%%%%%%%%%%%%%%%%%%%%%%%%%%%%%%%%%%%%%%%%
\wuhao\vspace*{2mm}

\begin{multicols}{2}

%%%%%%%%%%%%%%%%%%%%%%%%%%%%%%%%%%%%%%%%%%%%%%%%%%%%%%%%%%%%
%% Text of article.
%%%%%%%%%%%%%%%%%%%%%%%%%%%%%%%%%%%%%%%%%%%%%%%%%%%%%%%%%%%%
%    Section headings
\renewcommand{\baselinestretch}{1.08} \baselineskip 12.2pt\parindent=10.8pt

\renewcommand{\thefootnote}

\subsection{Current-sheet formation via MHD-thermodynamic interaction}

To complete the physical picture of spontaneous CS formation developed so far, we treat a field-fluid interaction in a low-$\beta$ environment.  This interaction was discovered theoretically in the investigation of the dynamic interiors of quiescent prominences [19,31,133--139].   Here it is presented as a general MHD process.

Consider the static equilibrium of a magnetic field ${\bm B}$ embedding in a tenuous plasma of density $\rho$ and pressure $p$, described by the equations:
\begin{align*}
\frac{1}{4 \uppi} \left( \nabla \times {\bm B}\right) \times {\bm B} - \nabla p -\rho g {\hat z}=\;&0, \tag {62}\\
\nabla \cdot \left[ \frac{\kappa}{|{\bm B}|^2} \left( {\bm B} \cdot \nabla T \right) {\bm B} \right] + {\mathcal R} - {\mathcal S} =\;& 0,\tag {63}
\end{align*}
where $g$ is the solar gravitational acceleration directed in the (Cartesian) vertical $-{\hat z}$ direction in the force-balance equation.  The other equation describes anisotropic thermal conduction with conductivity $\kappa$, directed everywhere along the magnetic field, its thermal flux balanced by radiative loss ${\mathcal R}$ and heat source ${\mathcal S}$ per unit volume.  We take $\kappa$, ${\mathcal R}$ and ${\mathcal S}$ to be explicitly known functions of $\rho$, temperature $T$ and the magnetic field ${\bm B}$.  Adopting the ideal gas law to relate $(p, \rho, T)$ and imposing the solenoidal condition (59) on ${\bm B}$, we then have a closed set of equations for the dependent variables $\left(p, \rho, T, {\bm B} \right)$.  Let us derive an interesting conclusion about this system, that CS formation is inevitable in the general solution involving a 3D magnetic field that is strong in the sense of $\beta \ll 1$.

The above problem assumes zero resistivity, $\eta=0$, and zero cross-field thermal conductivity, $\kappa_{\perp}=0$.  In the solar corona $(\eta, \kappa_{\perp})$ are small but not zero.  We are interested in situations characterized with a dimensionless constant $\epsilon=\frac{K_{\perp}}{\eta} \ll 1$ where we have replaced thermal conductivity $\kappa_{\perp}$ with thermometric diffusivity $K_{\perp}$, for a typical coronal density; $K_{\perp}$ and $\eta$ having the same physical dimension.  Consider electrical and thermal conduction due to the free electrons in fully ionized hydrogen.  The Spitzer plasma model [1] gives an estimated $\epsilon \approx 0.6~ \beta$, where the numerical coefficient $0.6$ is defined by atomic and thermodynamic constants [19].  For a low-$\beta$ plasma, cross-field thermal diffusion is significantly weaker than resistive field-diffusion.  In other words, as fluid and field gradients steepen monotonically, resistive effect becomes important before the cross-field thermal insulation breaks down.

Now consider the energy transport equation (63) under the assumption of $\eta=K_{\perp}=0$.
Each thin magnetic flux tube is insulated thermally from the adjacent flux tubes.
For a given total mass in a given flux tube, the Lorentz force has no component along it,
and $p$ and $\rho$ are related hydrostatically with the density scale height determined by the temperature $T$.
The profile of $T$ along the tube in a steady state must direct a field-aligned thermal conduction that brings
heat from over-heated regions, where $|{\mathcal S}|>|{\mathcal R}|$, to be radiated away in
regions where $|{\mathcal R}|>|{\mathcal S}|$.  Force and thermal balance along the field generally
cannot be maintained in the steady state for an arbitrarily prescribed total mass,
typically resulting in a thermal-gravitational collapse [136].
Even if such a collapse is avoidable, 3D fields of complex topology [11,15,23] produce temperature profiles along flux tubes that are generally discontinuous across the tubes, one tube thermally insulated from another. It follows that the fluid pressure is also discontinuous across the flux tubes. For force balance between adjacent tubes, a discontinuity in $p$ must be balanced by a compensating discontinuity in $B^2$. This is easily accommodated by the strong field except that a field discontinuity implies a CS that must dissipate resistively at a small but nonzero resistivity.

The condition $\epsilon={K_{\perp} \over \eta}\ll 1$ is crucial, indicating that the cross-field insulation as the principal cause of the discontinuous temperature can hold up to induce a CS to the point of its resistive dissipation.  As the CS dissipates, not only is there a change in field topology by magnetic reconnection but the flux tubes would also exchange mass.  The mass exchanges take place over entire magnetic flux surfaces in a chaotic manner, a process likely to render the total mass along each flux tube to be a discontinuous function over the flux tubes.  Noteworthy is the fact that this effect involves a weak CS, the strong field easily developing a small jump in $B^2$ to compensate for the fluid pressure jump.  Resistive dissipation via CSs is inevitably induced by a tenuous fluid, the more tenuous the fluid the greater the effect.   Thus such a fluid can flow readily across the strong field it embeds.

Recent unprecedented high-resolution observations from the {\it Hinode} Mission of Japan Aerospace Exploration
Agency and {\it Solar Dynamics Observatory} Mission of US National Aeronautics and Space Administration have revealed that the plasma interior of a prominence is dynamic on its small scales despite its stable macroscopic
appearance [103,104,133].  The cool prominence over these small scales take the form of a multitude of vertical narrow filaments that fall steadily at less than free fall speeds across their detected horizontal fields [31,133].
A global upward mass flux is ejected intermittently on the small scales everywhere from the relatively thin layer of partially-ionized atmospheric layer called the chromosphere, shown in the central subfigure in Figure 2 [140].  The bulk of this ejected cool mass heats up to coronal temperature and returns as condensing plasma everywhere [141,142].  This return flow may be the source of the falling vertical filaments in the prominence interior [31], a form of condensation due to the magnetic geometry of the prominence flux rope [109--112].  This cross-field drainage can cycle through a quiescent prominence an estimated order of magnitude more mass in a day than the total mass maintained quasi-steadily in the prominence over days to a week [134].

To apply to the drainage in a real prominence, we need physically more complete models that account
for the global field in realistic 3D geometry, the partially-ionized state of the prominence,
the fully ionized corona, and other relevant physical features.  The above analysis based
on $\epsilon={K_{\perp} \over \eta}\ll 1$ serves only as a simple demonstration of an
MHD-thermodynamic effect. This general effect is physically distinct from the Parker spontaneous
CS formation but the two effects share a common feature. Each effect arises from the demand of
force (and thermal) balance imposed at each point in space subject to a global constraint,
a given total mass to be thermally distributed along a flux tube in one case and the
given field topologies invariant along three interacting flux tubes in the other case.
Generally this demand can be met only in a discontinuous field containing TDs under
the condition $\eta=0$. A similar field steepening to
form TDs can thus be expected in a tenuous ($\beta\ll 1$) fluid when the condition
$\eta=0$ fails weakly, except that unchecked field
steepening results in resistive diffusion of fields at dynamically relevant time scales.

\section{Summary}

Our review begins with magnetic flux conservation as the defining property of ideal induction equation
(5).  This property is Lagrangian in nature, to be observed only by following a specific
fluid surface as it deforms continuously in the flow.  The exhaustive partition of the fluid into
disjoint, unlinked, thin $\tau$ toroids sets the stage for just such an observation.
Each $\tau$ toroid sees the given field in terms of two conserved fluxes, its axial
flux $f(\tau)$ and the flux $F(\tau)$ trapped in the hole of the toroid.
The solenoidal condition is the reason the given field is made up of only two
independent flux systems, seen in the fact that a solenoidal vector field is
defined by two free scalar functions in 3D space, the CK field representation
a case in point.  The general Lagrangian helicity ${\mathcal H}$ defined by eq. (18) then follows naturally.
In its simplest form, the Lagrangian helicity $H_{\rm L}$ is the sum of products of
the two fluxes $f(\tau)$ and $F(\tau,\upsilon)$ associated with each toroid, a measure of entanglement between the two flux systems.

Historically magnetic helicity had been formulated in the Eulerian description in common use:
the classical total helicity $H_{\rm c}$ of a contained field and the relative total
helicity $H_{\rm R}$ of an anchored field, the latter a relative measure against an
associated potential field as a reference. The discovery of the total absolute helicity $H_{\rm abs}$
dispenses with the need for a reference field, describing the field entanglement in both
contained and anchored fields on an equal conceptual basis.  The Eulerian total helicity,
defined in the different ways, is just one of a continuum of constants of motion describing
the frozen-in field topology under perfect electrical conductivity.
The Lagrangian continuum of conserved ${\mathcal H}$ suggests that corresponding to
it must exist a continuum of general Eulerian
helicity ${\mathcal H}^*$ that awaits discovery.

The general MHD ideas discussed in the review are given specificity by the cylindrical domain $V$.
The expression of the total absolute helicity $H_{\rm abs}$ in this domain depends on the use of a
specialized CK field representation in terms of $\left[ {\bm B}_{\Psi}, {\bm B}_{\Phi} \right]$,
subject to boundary conditions (30)--(32) on the two generating functions $(\Psi, \Phi)$.
This specialized definition of $H_{\rm abs}$ leaves open for future development as to how $H_{\rm abs}$ may be defined for general domains without the facility of a CK representation.  The question can be posed differently: how may the CK representation be generalized for a field in a domain of an arbitrary shape?

We have centered our discussion on the cylindrical $V$ for a reason worth pointing out.
In the two other special domains admitting CK representations, namely, the finite space
between two concentric spheres and its limiting case of the unbounded space between two
parallel planes, the total absolute helicity $H_{\rm abs}$ and the relative total
helcity $H_{\rm R}$ have the same value.  A topological/physical
understanding of why these two domains have this special property remains to be found.  All three helicities $[H_{\rm R}, H_{\rm abs}, H_{\rm L}]$ are typically distinct [32,33,35] in $V$, likely to be representative of the general domain.

The CK field representation in $V$ is a physical realization of the Lagrangian $\tau$-partition of a fluid.  It can initiate a Lagrangian description by the $\tau$ partition defined at an initial time.  Alternatively, as an Eulerian description, the CK representation decomposes the given field into two linearly superposed fields at each moment in time without reference to where each fluid parcel has moved.

The Lagrangian description seems conceptually simpler, although computationally complicated.
The CK representation at any given time partitions the given field into the two fields $\left[ {\bm B}_{\Psi}, {\bm B}_{\Phi} \right]$ that subsequently independently evolve in time with the common fluid velocity ${\bm v}$.  At any subsequent time, the two fields are greatly deformed, depending on ${\bm v}$, but their linear sum always defining the given field at each moment.  Such a decomposition has value only if the two components in the superposition are simple fields in terms of which the given field in its admissible complexity can be expressed.  The relationship between the CK and Euler-potential representations shows that such a simple field may be defined to be one that can be represented by a pair of globally defined Euler potentials.  Then all fields can be expressed as the sum of up to three simple fields.  Sect. 2.5 on vortex dynamics illustrates these field representations in instructive contrast to ideal MHD. Thus we have a physically complete description of the frozen-in field topology in ideal MHD.

The Parker theorem was discovered in the theoretical investigation of the solar corona [7,8,29].  A large volume of work has been published over many years and the reader is referred to recent comprehensive reviews [10,17,22].  The basic point is that a continuous field is generally incompatible with force-free equilibrium at each point in space if the field is arbitrarily prescribed with a complex 3D topology.  The nature of the equilibrium at each point in space is not crucial.  The incompatibility arises between the point-by-point and global conditions, the former described by PDEs and the latter by integral equations [17].  For example, a similar incompatibility may be found in a steady field-aligned flow if the field topology is to be arbitrarily prescribed [143--145].

The partition of the space ${\mathcal B}$ of all continuous fields in $V$ into the disjoint subspaces ${\mathcal B}_T$, each comprising fields with a common topology $T$, serves well to describe the theorem.  The mathematical structure of subspace ${\mathcal B}_T$ is defined by the shape of the general field domain $V$, taken to be an upright cylinder for specificity in our review, and field topology $T$. The continuous force-free fields, if any, belonging to ${\mathcal B}_T$ are $\delta E_{\rm M}=0$ extremum points in that function space. The subset of these extremum points corresponding to $E_{\rm M}$ being a local minimum are the possible end-states terminating the viscous relaxation of a field with topology $T$.  In addition to these minimum-$E_{\rm M}$ points belonging to ${\mathcal B}_T$ are possible minimum-$E_{\rm M}$ points representing force-free fields with equilibrium TDs also of topology $T$.  These discontinuous force-free fields do not belonging to ${\mathcal B}_T$, of course, but each may be the limit point of a path of viscous relaxation in ${\mathcal B}_T$, an intuitive picture of what is meant by the space ${\mathcal B}_T$ not being Cauchy complete.  There is a richness in the structure of ${\mathcal B}_T$ not trivial to construct, but this picture shows the way one might build explicit models to demonstrate the Parker theorem.

One possibility is to discover a field domain and a field topology $T$ such that ${\mathcal B}_T$ is demonstrably without a minimum-$E_{\rm M}$ point contained in it. Then all paths of viscous relation in ${\mathcal B}_T$ must lead to a minimum-$E_{\rm M}$ force-free field with TDs. In such a case, ${\mathcal B}_T$ may have one or more $\delta E_{\rm M}=0$ extremum points except that none of them are local minimum in $E_{\rm M}$. These extremum points describe continuous force-free fields that are linearly unstable.  When perturbed, the field evolves away from its unstable initial equilibrium with an inevitability of forming TDs.  The first demonstration [146] of the Parker theorem of this kind has opened up a promising approach of inquiry, investigating the structures of all vector function spaces ${\mathcal B}_T$.

The dependence of ${\mathcal B}_T$ on the shape of the field domain presents a different aspect of the Parker theorem.   The reader is referred to interesting results published on the availability of a continuous force-free state following a simultaneous, continuous deformation of the field and its domain [12,17,60--62,81,90,91].

The spontaneous formation of CS is also encountered in the thermal and force balance of a radiating heated fluid subject to an anisotropic thermal conduction strictly channelled along a strong frozen-in magnetic field.  Here a continuous field is incompatible with thermal and force balance at each point in space if the total mass loaded on each flux tube is arbitrarily prescribed [136]. This CS formation is physically distinct from that described by the Parker theorem, the former a consequence of thermodynamics and the latter field topology, the two effects expected to be acting simultaneously in a real situation.

Gravity plays an important role in the static MHD-thermodynamic coupling treated.  The gravitational settling of a heated radiating fluid by the formation of CSs necessarily involves the drainage of the fluid across the field supporting its weight.  In the case of a quiescent prominence, this drainage has been estimated from observation to be significant [134].  The formation of an interstellar cloud supported by the galactic magnetic field is a similar MHD process [147--154].  The spontaneous formation of CSs via this MHD-thermodynamic coupling is a promising mechanism for an MHD fluid to separate from, or to flow across [31], its embedding fluid in spite of a weak resistivity, an issue important to the current understanding of star-formation out of a magnetized plasma.

The theoretical developments reviewed are motivated by the observed solar corona.
This hydromagnetic atmosphere is almost a perfect electrical conductor at its million-degree temperature.
The ubiquitous heating of the corona maintaining its temperature with stability and
the rapid liberation of dissipated energies in violent flares may be explained in terms of spontaneous
CSs in the rich variety of physical circumstances represented by these observed phenomena.
CS formation is due to high conductivity coupled to the dynamical forces, but this same coupling
also sets macroscopic constraints on the amount of magnetic energy possible to
liberate via this ubiquitous process in the solar corona. The Taylor hypothesis is the simplest
form of such constraints that produce macroscopic magnetic structures.
This self-organization in turbulent MHD is a possible energy-storage mechanism that fuel flares and coronal mass ejections.

There has been considerable progress in the basic MHD theory reviewed, leading to clarification of basic concepts and clearly articulated problems for research.  Among these problems, the most pressing might be the formulation of a general theory for the breakdown of flux conservation in the low-resistivity limit, going beyond the Taylor hypothesis.  Basic MHD theory is essential for interpreting numerical simulations as well as the phenomena observed in the solar corona [155--158].

\vspace*{2mm} \Acknowledgements{\bahao I thank Profs. CHEN PengFei, JUDGE Phil  and PARKER Gene  for critical comments on the manuscript.  The National Center for Atmospheric Research is sponsored by the US National Science Foundation.}

\end{multicols}

\end{document}